\documentclass[twocolumn]{aastex62}

\usepackage{epstopdf}
\usepackage{amsmath}
\usepackage{soul}  

\received{---}
\revised{---}
\accepted{---}
\submitjournal{AJ}

\begin{document}
\title{\uppercase {Shepherding in a Self-gravitating Disk of Trans-Neptunian Objects}  }
\shorttitle{Shepherding of trans-Neptunian objects}

\author[0000-0003-4623-1165]{Antranik A. Sefilian}
\email{aas79@damtp.cam.ac.uk}
\affil{Department of Applied Mathematics and Theoretical Physics, University of Cambridge,\\
Centre for Mathematical Sciences, Wilberforce Road, Cambridge CB3 0WA, UK }

\author{Jihad R. Touma}
\email{jt00@aub.edu.lb}
\affiliation{Department of Physics, American University of Beirut,\\
PO BOX 11-0236, Riad El-Solh, Beirut 11097 2020, Lebanon}

\shortauthors{Sefilian \& Touma}

\begin{abstract}

A relatively massive and moderately eccentric disk of trans-Neptunian objects (TNOs) can effectively counteract apse precession induced by the outer planets, and in the process shepherd highly eccentric members of its population into nearly-stationary configurations which are anti-aligned with the disk itself. We were sufficiently intrigued by this remarkable feature to embark on an extensive exploration of the full spatial dynamics sustained by the combined action of giant planets and a massive trans-Neptunian debris disk. In the process, we identified ranges of disk mass, eccentricity and precession rate which allow apse-clustered populations that faithfully reproduce key orbital properties of the much discussed TNO population. The shepherding disk hypothesis is to be sure complementary to any potential ninth member of the Solar System pantheon, and could obviate the need for it altogether. We discuss its essential ingredients in the context of Solar System formation and evolution, and argue for their naturalness in view of the growing body of observational and theoretical knowledge about self-gravitating disks around massive bodies, extra-solar debris disks included. 

\end{abstract}


\section{Introduction} \label{sec:intro}

Trans-Neptunian (phase)-space appears to be populated with bodies that show signs of orbital sculpting, then shepherding. With the discovery of $2012 ~ \text{VP}_{113}$, a Sedna-like object, \citet{tru14} first argued for a ninth planet of 5 $M_{\earth}$ on a circular orbit at 200 AU as a potential shepherd of several TNOs with eccentric and inclined orbits showing peculiar clustering in the argument of periapse. Later on, \citet{bat16} noted remarkable spatial nodal alignment of the same objects. They reexamined the proposition of an additional planet, and argued instead for a super-Earth (dubbed ``Planet Nine") on a larger eccentric and inclined orbit, while  appealing to an alternative resonant process for the aligning trap \citep{bat16}. Further indirect evidence for such a planet was sought around apparent deviations in the orbit of the Cassini spacecraft \citep{fie16}, and in the potential to explain the Sun's obliquity \citep{bai16,lai16}. 

To date, twenty-three trans-Neptunian objects (TNOs) have been identified on eccentric and inclined orbits, with semi-major axes $a_p\succsim150$ AU and perihelion distance $q_p >30$ AU. Out of these, thirteen roam with $a_p \succsim 250$AU and have had their notorious kinematic properties classified in the course of Planet-Nine related studies which propose to explain them. They are interpreted as either: spatially clustered and anti-aligned with Planet Nine (ten objects); spatially clustered and aligned with Planet Nine (two objects); neither here nor there, though strongly perturbed by Planet Nine (one object). These classes are of course expected to grow in size and definition by proponents of a ninth planet which is requested to structure, along with the rest of the Solar system, the phase space in which TNOs are presumed to evolve.

Alternatively, \citet{sha17a} argued that the spatial clustering which Planet Nine is supposed to explain is fraught with observational bias. Running their own orbital simulations, they disputed the claim that a planet alone could maintain clustering for the required duration. They further noted that to observe this group of TNOs within existing campaigns, implies a parent population of $6-24 \, M_{\earth}$. Such a massive reservoir of trans-Neptunian icy bodies is nearly two orders of magnitude larger than currently favored estimates \citep{gla11}.  \citet{sha17a} took this requirement as further evidence against significant clustering, and gave no further consideration to the dynamical signature of a massive trans-Neptunian population. 

Here, we go precisely after the dynamical impact of an extended and relatively massive disk of trans-Neptunian objects, and demonstrate that it alone can provide a fair amount of  shepherding, perhaps obviating the need for an extra planetary member in the Solar System pantheon, surely complementing it. 

We describe results in a progression of complexity around a fiducial razor thin disk. We then comment briefly on parametric variations on such a disk, discuss its properties and their origin, together with the potential interplay between the dynamical features it stimulates, and those associated with a hypothetical few Earth mass planet in post-Neptunian realm.

\section{Coplanar Dynamics} \label{section:fiducial}

We study the secular, orbit-averaged coplanar dynamics of trans-Neptunian test particles characterized by their semi-major axis $a_p$, eccentricity $e_p$ and apsidal angle $\varpi_p$, in the combined gravitational potential of: {\bf a}- the outer planets  and {\bf b}- a hypothetical extended disk, lying in the plane of the giant planets, and built out of confocal eccentric apse-aligned orbits.

The outer planets are included via the quadrupolar potential of a sequence of fixed concentric circular rings. The coplanar disk is parametrized by its non-axisymmetric surface density $\Sigma$ (Eq. \ref{eqn:A1}), eccentricity profile $e_d$,  global apsidal angle $\varpi_d$ (fixed at $\pi$, except otherwise stated), and inner and outer boundaries $a_{in}$ and $a_{out}$ respectively.

We work with disks that have power-law density/eccentricity profiles
\begin{equation}
\Sigma_d(a_d) = \Sigma_0 \bigg( \frac{a_{out}}{a_d}   \bigg) ^p 
\label{eqn:Sigma_d}
\end{equation}
and
\begin{equation}
e_d(a_d) = e_0 \bigg( \frac{a_{out}}{a_d} \bigg)^q
\label{eqn:e_d}
\end{equation}
for $a_{in} \leq a_d \leq a_{out}$. Here, $\Sigma_0$ and $e_0$ are the pericentric surface density and eccentricity at the outer edge of the disk respectively. Surface density profiles with  $p < 2$ ($p > 2$) are associated with disks which have more mass concentrated in the outer (inner) parts of the disk than in the inner (outer) regions. Total disk mass $M_d$ can be estimated with $M_d \simeq 2\pi \int_{a_{in}}^{a_{out}} \Sigma_d(a_d) a_d da_d$ yielding
\begin{equation}
M_d = \frac{2\pi}{2-p} \Sigma_0 a_{out}^2 \bigg[  1 - \bigg( \frac{a_{in}}{a_{out}} \bigg)^{2-p}   \bigg]  
\approx \frac{2\pi}{2-p} \Sigma_0 a_{out}^2
\label{eqn:Md}
\end{equation}
where the approximation is valid as long as the disk edges are well-separated and more mass is found in the outer parts. Disk models which were thoroughly explored in this work are listed in Table \ref{tab:tab1}. 
\begin{table}[h!]
\begin{center}
\caption{Power-law disk models. \label{tab:tab1}}
\begin{tabular}{cccccccc}
\tableline
\tableline
Disk Model & p & q & $a_{in}$ & $a_{out}$ & $\varpi_d$ & $e_0$ & $M_d$ \\
&  &  & (AU) &(AU) & (rd) &  & $(M_{\earth})$ \\
\tableline
DM1 & 0.5 &-1 & 40  & 750 &$\pi$ & 0.165 & 10 \\
DM2 & 0.5 &-1 & 40  & 750 &$\pi$ & 0.165 & 2.5 \\
DM3 & 0.5 &-1 & 40  & 750 &$\pi$ & 0.165 & 20 \\
DM4 & 2.5 &-1 & 40  & 750 &$\pi$ & 0.165 & 10 \\
\tableline
\end{tabular}
\tablecomments{Disk Model 1 (DM1) is the fiducial disk configuration adopted in this work.}
\end{center}
\end{table}

Our basic shepherding mechanism is best articulated in planar dynamics which will ultimately provide the skeleton around which fully inclined behavior is structured (see Section \ref{section:3}).

At the outset it is important to remind the reader that hot nearly-Keplerian disks induce negative apse precession in their constitutive particles, in contrast to the familiar prograde apse precession expected from cold disks of isolated planets. This fact was recently noted to argue for the role of massive gaseous disks in mitigating the destructive role of perturbations induced on planetesimal disks by wide binary companions \citep{raf13, sil14,raf14,sef17}.

We exploit that feature here by appealing to the negative precession induced by an extended and massive trans-Neptunian debris disk to mitigate against and, if possible, freeze the prograde differential precession induced by the outer planets on a distinguished population of TNOs which is yet to be identified. 

With this in mind, we recover the secular orbit-averaged disturbing potential, $R_d$, of power law disks up to fourth order in the orbital eccentricity of a coplanar test particle (see Appendix \ref{appendix1}): 
\begin{eqnarray}
R_{d} &=& K \bigg[ \psi_1 e_p \cos\Delta \varpi + \psi_2 e_p^2 + \psi_3 e_p^2 \cos(2 \Delta \varpi) \nonumber 
\\&+& \psi_4 e_p^3 \cos \Delta \varpi + \psi_5 e_p^4         \bigg],
\label{eqn:Rd}
\end{eqnarray} where
\begin{subequations}
\begin{align}
& K \equiv \pi G \Sigma_0 a_{out}^p a_p^{1-p} > 0, 
\label{eqn:K}
\\
& \Delta \varpi \equiv \varpi_p - \varpi_d.
\end{align}
\end{subequations}
The dimensionless coefficients $\psi_{i}$ are given by equations \ref{eqn:psi1} -- \ref{eqn:psi5}.

Orbit averaged quadrupolar action of the outer planets is captured via
\begin{equation}
R_{p} = + \frac{1}{3} \Gamma (1-e_p^2)^{-\frac{3}{2}},
\label{eqn:outer-quad}
\end{equation}
with 
\begin{equation}
\Gamma =  \frac{3}{4} \frac{G M_{\odot}}{a_p} \sum_{i = 1}^{4} \frac{m_i a_i^2}{M_{\odot}a_p^2},
\label{eqn:GammaHp}
\end{equation}
and $[(m_i, a_i), i = 1..4]$ the masses and current semi major axes of the four giant planets. 

Combining both contributions, Hamilton's equations for the signed ``angular" momentum  $l_p = \pm \sqrt{1-e_p^2} $ and the conjugate longitude of the apse  $\varpi_p$ are given by:
\begin{eqnarray}
L_p\,\dot{l}_p &=& - K \bigg[\sin\Delta\varpi \big[ \psi_1 \sqrt{1-l_p^2} + \psi_4 (1-l_p^2)^{\frac{3}{2}}  \big] \nonumber 
\\
&+& 2  \psi_3 (1-l_p^2)   \sin (2\Delta\varpi)   \bigg],
\label{eqn:ldot}
\end{eqnarray}
and
\\
\begin{eqnarray}
L_p\,\dot{\varpi}_p  &=& \frac{\Gamma}{l_p^4} + Kl_p \bigg[ \frac{\psi_1\cos\Delta\varpi}{\sqrt{1-l_p^2}} + 2\psi_2 + 2\psi_3\cos (2\Delta\varpi)  \nonumber \\
&+& 3\psi_4 \cos\Delta\varpi \sqrt{1-l_p^2} + 4\psi_5(1-l_p^2)           \bigg],
\label{eqn:wdot}
\end{eqnarray}
with $L_p = \sqrt{G M_{\odot} a_p}$, the constant angular momentum conjugate to the mean anomaly which has been averaged out of the game. {Disturbing functions [Eqs.(\ref{eqn:Rd}, \ref{eqn:outer-quad})], and equations of motion [Eqs.(\ref{eqn:ldot}, \ref{eqn:wdot})] govern the dynamics of both prograde and retrograde orbits which are coplanar with disk and planets. Below, and in keeping with observed aligned TNOs, we concentrate primarily on the prograde phase space. }

\begin{figure}
\epsscale{1.2}
\plotone{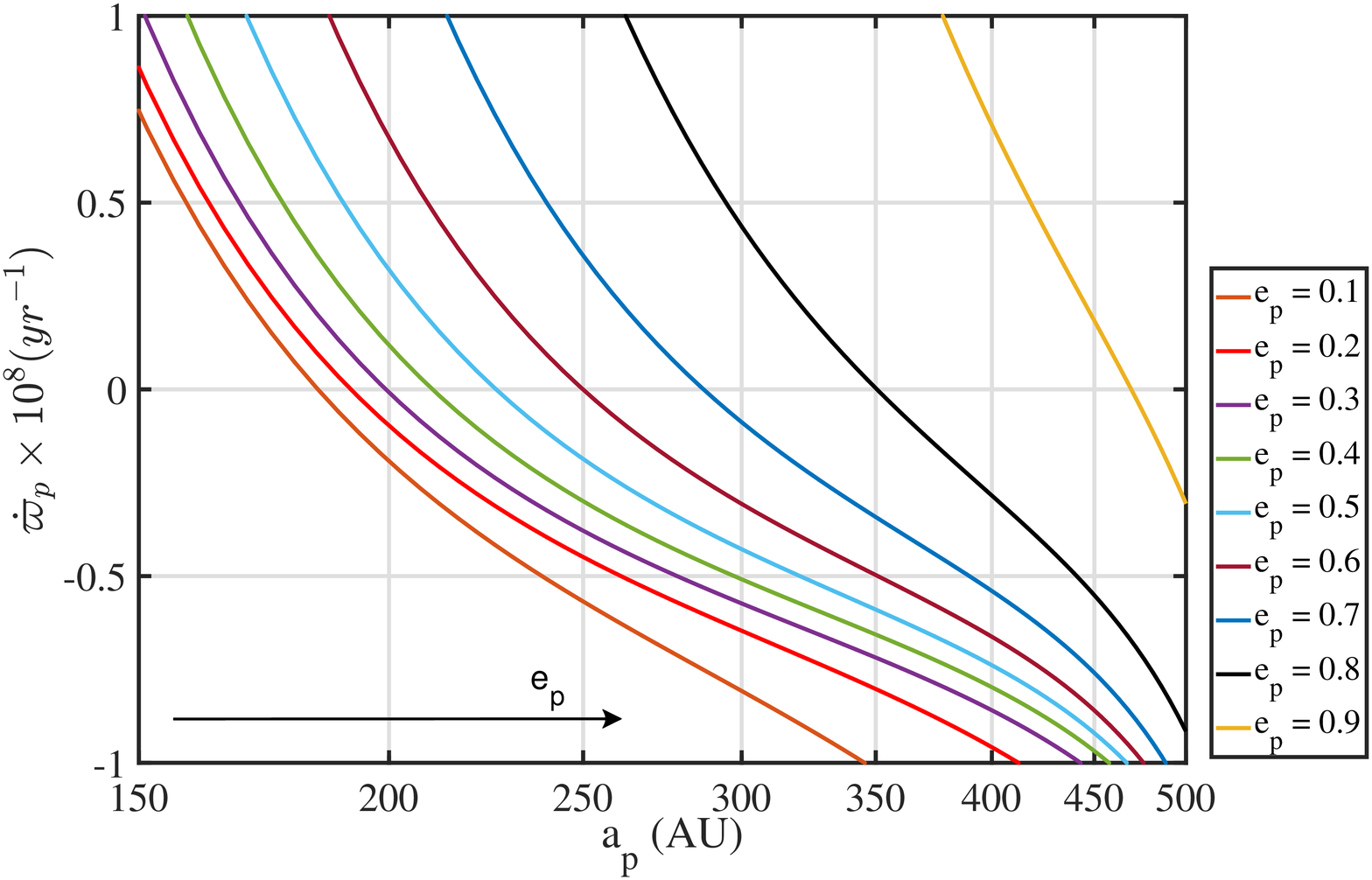}
\caption{Rate of apse precession $\dot{\varpi}_p$  of TNOs which are anti-aligned with the disk ($\Delta \varpi = \pi$), over a range of semi-major axis $a_p$ for different values of eccentricity $e_p$.  Precession is here driven by the combined action of the giant planets and the fiducial disk model DM1 (see Table \ref{tab:tab1}). Zero apse precession obtains for all the considered values of $e_p$, and semi-major axes $a_p$ between 150 and 500 AU. Given that the torque vanishes for $\Delta\varpi=\pi$ (see Eq. \ref{eqn:ldot}), we have here a family of stationary orbits, which are anti-aligned with the disk, and whose eccentricity grows with $a_p$.
\label{fig:fig1}}
\end{figure}

In Figure \ref{fig:fig1}, we display the apsidal precession rate induced by the outer planets and the fiducial power law disk model (Table \ref{tab:tab1}, model DM1) on orbits which are anti-aligned with the disk's spatial orientation (i.e orbits with $\Delta \varpi = \pi$), over a range of semi-major axis $a_p$, and for different values of TNO eccentricity $e_p$. Evidently, there is an eccentric anti-aligned orbit with zero net apse precession at all semi-major axes in the considered range. Keeping in mind that the torque (Eq. \ref{eqn:ldot}) is null for $\Delta \varpi = \pi$, we have here evidence for a one parameter family of anti-aligned stationary orbits which will provide the skeletal structure around which the observed TNOs, and the rest of our paper will be fleshed out! 

\begin{figure}
\epsscale{1.2}
\plotone{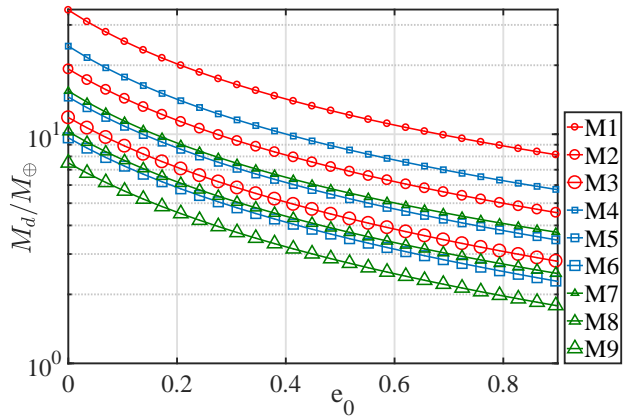}
\caption{The required disk mass $M_d$ as a function of disk eccentricity ($q = 0$, $e_d(a_p) = e_0$) to obtain stationary anti-aligned  ($\Delta \varpi = \pi$) TNO orbits at $a_p = 257$ AU with $e_p = 0.82$. The calculation is performed for disk parameters ($p,~a_{in} ~\text{and}~ a_{out}$) given in Table \ref{tab:tab2} . When combined with results shown in Figure \ref{fig:fig1}, this figure speaks for the robustness of our proposed mechanism. 
\label{fig:fig2}}
\end{figure}

Before we examine the full dynamical behavior of this family,  we thought it reasonable to probe the robustness of this remarkable broad-ranged cancellation of apse-precession to variations in disk properties (mass density profile, disk eccentricity, disk radial extent). We thus computed the disk mass $M_d$ which is required to apse-freeze an anti-aligned orbit ($\Delta\varpi = \pi$) of given eccentricity $e_p$ and semi-major axis $a_p$ when embedded in a disk of given  mass distribution (dictated by $p$), inner and outer edge, and $e_0$. The outcome of this exercise for  a test particle with $a_p = 257$ AU and $e_p = 0.82$ is shown in  Figure \ref{fig:fig2}, and permits the following conclusions: {\bf a-} the required disk mass can be as low as $\sim 1 M_{\earth}$ and as high as $\sim 30 M_{\earth}$; {\bf b-} lower $M_d$ is required at higher disk eccentricity, an effect which is surely due to enhancement of disk induced retrograde precession with increasing $e_0$; {\bf c-} the critical $M_d$ increases with increasing $a_{out}$: this behavior is evident in axisymmetric disks where the disk induced precession is well approximated by the following expression\footnote{The approximate expression of $\dot{\varpi}_p \big|_{disk}$ is obtained for circular disks with $p=1$ under the reasonable assumption of ${a_{in} \ll a_p \ll a_{out}}$. This assumption allows us to drop contributions to $R_d$ (Eq. \ref{eqn:Rd}) from the disk edges rendering the coefficients $\psi_i$ mild functions of only $p$ and $q$ (see Eq. \ref{eq:psi1_expanded}-\ref{eq:psi5_expanded}). {For instance, in a circular disk $\psi_2 = -0.5$ for p = 1; } see Eq. \ref{eq:psi2_expanded}.}
\begin{equation}
\dot{\varpi}_p \big|_{disk} \simeq -4.2\times 10^{-10} \text{yr}^{-1} \frac{M_d}{ 1M_{\earth}}  \frac{10^3AU}{a_{out}} \bigg(\frac{a_p}{500AU}\bigg)^{-0.5} 
\label{eqn:wdot_approximation}
\end{equation}
for circular TNO orbits.
What is evident for axisymmetric disks is clearly maintained in eccentric ones. Furthermore, we checked that our conclusions for a single anti-aligned equilibrium orbit (with $e_p = 0.82$ and $a_p = 257$ AU) holds for all. 
\begin{table}
\begin{center}
\caption{Power-law disks used to generate Figure \ref{fig:fig2}. \label{tab:tab2}}
\begin{tabular}{cccc}
\tableline
\tableline
Model & $a_{in}$(AU) & $a_{out}$(AU) & p
 \\
\tableline
M1 & 200 & 1200 & 0.1  \\
M2 &  & & 0.5  \\
M3 &  & & 0.9  \\
M4 & 200 & 1000 & 0.1  \\
M5 &  & & 0.5  \\
M6 &  & & 0.9  \\
M7 & 200 & 800 & 0.1  \\
M8 &  & & 0.5  \\
M9 &  & & 0.9  \\
\tableline
\end{tabular}
\tablecomments{We have adopted a constant disc eccentricity by setting $q=0$, with $0 \leq e_d(a_p) =  e_0 \precsim 0.90$.}
\end{center}
\end{table}

\begin{figure}
\epsscale{1.2}
\plotone{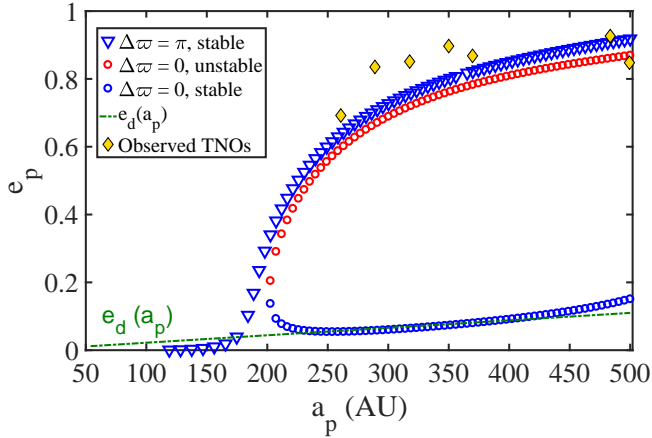}
\caption{The stationary TNO families ($e_p, \Delta \varpi$), both stable and unstable, which are sustained by DM1 (Table \ref{tab:tab1}) acting together with the giant planets. The stable anti-aligned ($\Delta\varpi = \pi$) family follows quite closely the observed $e_p - a_p$ trend of the seven clustered TNOs which are considered in this study (Table \ref{tab:tab3}). A stable family of non-precessing orbits which are aligned with the disk ($\Delta\varpi = 0$) has an eccentricity profile which is almost identical to the imposed disk eccentricity profile $e_d(a_p)$.
\label{fig:fig3} }
\end{figure}

Exhaustive exploration of the dynamics sustained by our orbit averaged Hamiltonian shows that the fiducial disk model (Table \ref{tab:tab1}, DM1) harbours three distinct families of orbits:
\begin{itemize}
\item A family of stable, highly eccentric, and anti-aligned orbits ($\Delta \varpi = \pi$): this family shows equilibrium $e_p(a_p)$-behavior which is remarkably consistent with the trend followed by clustered TNOs. It is the family of most interest to us in relation to the shepherding phenomenon.
\item  A family of stable aligned ($\Delta \varpi = 0$) and low eccentricity orbits: interestingly enough, this family follows in its trend the eccentricity distribution of the disk that hosts it.
\item A family of highly eccentric and aligned orbits ($\Delta \varpi = 0$): this family parallels the behavior of the stable high $e_p$ anti-aligned family but is doomed to instability. 
\end{itemize}
Taking it for granted that the stable anti-aligned family correlates with the observed family of clustered TNOs, we conclude that DM1 naturally excludes stable high eccentricity clustering in the opposite apse orientation, an orientation where significant high eccentricity clustering is apparently not observed\footnote{We know of two highly eccentric TNOs [$2013 ~ \text{FT}_{28}$ \citep{she16} and $2015 ~ \text{KG}_{163}$ \citep{sha17b}] having, $a_p > 250$ AU and $q_p > 30$ AU, and which are currently anti-aligned with the much discussed clustered bunch. Their dynamical behavior is reviewed in Section \ref{discussion}.}. All three families are shown in Figure \ref{fig:fig3} together with the eccentricity distribution of the underlying disk. 

\begin{figure*}
\gridline{\fig{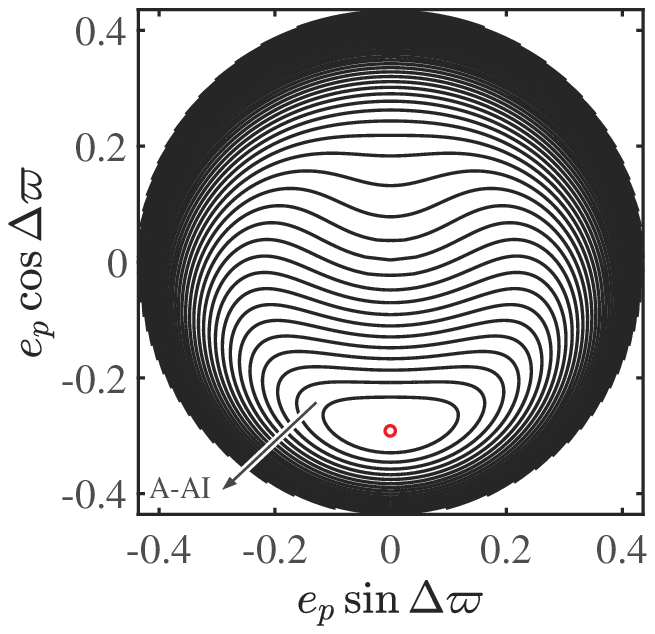}{0.3\textwidth}{(a) $a_p = 198$ AU}
          \fig{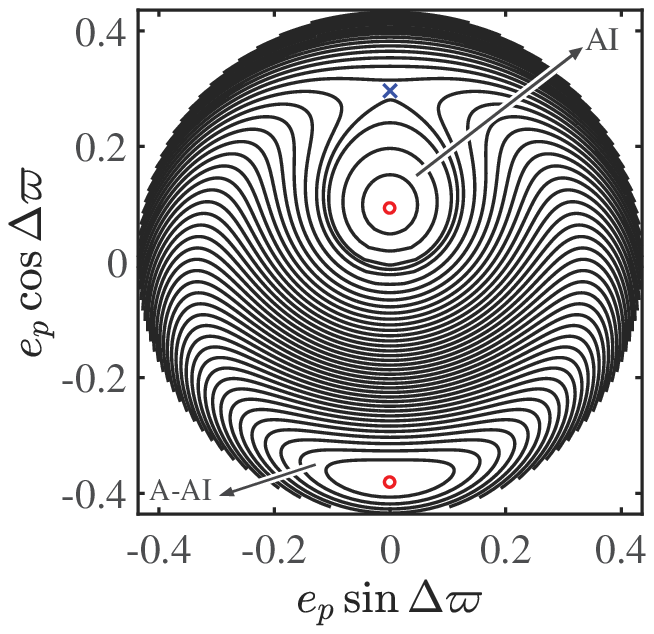}{0.3\textwidth}{(b) $a_p = 207$ AU}
          \fig{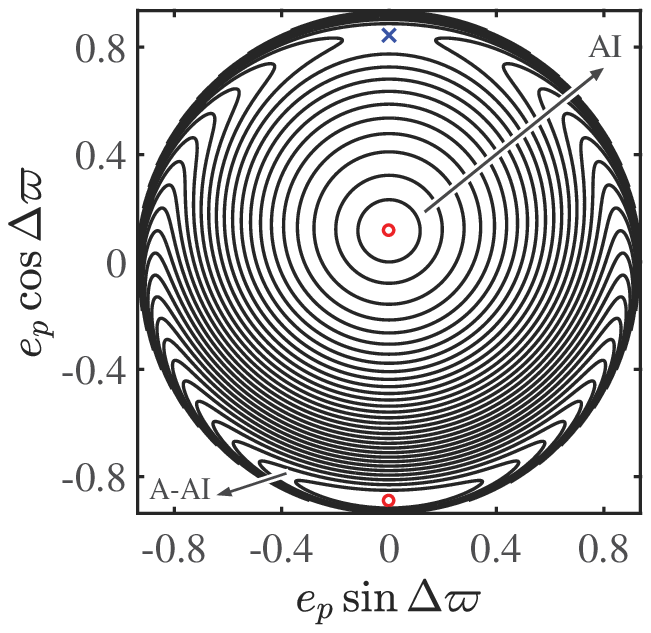}{0.3\textwidth}{(c) $a_p = 453$ AU}}
\caption{Phase portraits corresponding to the total Hamiltonian, $H_p + H_d = - (R_p + R_d)$, in $e_p (\sin \Delta\varpi, \cos\Delta\varpi)$ space, at three different TNO semi-major axes $a_p$ for the fiducial disk model DM1. The semi-major axes were chosen to illustrate the existence and bifurcation of the families identified in Figure \ref{fig:fig3}. The stable (unstable) secular equilibria are highlighted in red (blue). Panel {\bf a} shows the phase portrait at $a_p=198$ AU with a single stable anti-aligned equilibrium and its associated Anti-Aligned Island (A-AI) situated at $\Delta\varpi = \pi$. In Panels {\bf b} and {\bf c}, we show the phase portrait at $a_p= 207$ AU and  $a_p=453$ AU respectively, with two new aligned equilibria ($\Delta\varpi = 0$), one unstable and one stable, the latter coming with an Aligned Island (AI) of librating orbits. The $e_p-a_p$ trends of Fig. \ref{fig:fig3} are evident with the progression in semi-major axis, through panels {\bf a},  {\bf b} and {\bf c} respectively. 
\label{fig:fig4}}
\end{figure*}

These families can be further situated within the global phase space structure which is captured in Fig. \ref{fig:fig4} at three distinct semi-major axes. In addition to equilibria and their bifurcations, the phase diagrams reveal aligned and anti-aligned islands (AI and A-AI respectively) of bounded motion around the parent stable equilibrium orbits. These islands host orbits which will show signs of clustering (aligned and anti-aligned respectively) when considered collectively, and in time. The A-AI shelters high eccentricity orbits which straddle, as they oscillate in their eccentricity and longitude of apse, the parent equilibrium family which so closely follows the $e_p-a_p$ trend of the observed TNOs. We will have more to say about this population when we discuss it in full 3D glory below. The AI on the other hand is populated by orbits which share, on average,  the orientation and orbital eccentricity of the host disk, thus providing a rich supply of orbits with which to construct  a self-consistent deformation of DM1 (an exercise on which we comment in Section \ref{discussion}). It further includes orbits which have large amplitude eccentricity variations that bring them close to the unstable high eccentricity aligned orbits. Such orbits tend to linger around that unstable aligned configuration,  projecting a temporary sense of eccentric alignment with the disk which is then lost to evolution on timescales which are long enough\footnote{Similar such orbits which tend to linger around eccentric aligned orientation can be found in the A-AI, when one moves sufficiently far from the equilibrium. These orbits are fairly eccentric, with some suffering encounters with Neptune on their journey.}. 

The AI and A-AI are both surrounded by high eccentricity orbits which circulate in the longitude of the apse, while maintaining large and near constant eccentricity. More on these populations when we discuss curious members of the TNO population in Section \ref{discussion}. 


In sum, DM1 shepherds eccentric anti-aligned orbits ($\Delta\varpi = \pi$) whose properties favor them as coplanar analogs of the family identified by \citet{tru14}, while at the same time supporting aligned and non-precessing orbits of moderate eccentricity which promise to reproduce the disk that supports them, in a self-consistent treatment of the dynamics\footnote{The coplanar dynamics we just mapped out is naturally in dialog with \citet{beu16} who considers secular dynamics which is controlled by the outer planets together with a coplanar, eccentric, Planet Nine like object. He identifies eccentric non-aligned secular equilibria which are analogous to the ones we recover here. Perhaps, the analogy can be pushed further to argue for the combined action of a pre-existing disk and a scattered planet. We discuss this possibility, but do not explore its detailed workings in the present article.}. It would thus seem that a massive eccentric trans-Neptunian debris disk together with the action of the outer planets provides significant and profoundly suggestive clustering of embedded test particles. Whether such a disk obviates the need for a Planet Nine-like perturber altogether will be discussed further below after we explore out-of-plane dynamics, close to where the observed TNOs tend to roam. However, what is already clear at this coplanar stage is that the action of such a potential disk (which is evidently felt by highly eccentric orbits for disks with mass as low as $\sim 1~ M_{\earth}$) cannot be ignored, and will have to be considered together with any putative extra planet. 

\section{Life outside the Plane: Going 3D} \label{section:3}

Freezing coplanar orbits is interesting enough. However, the observed TNO bunch is held together in inclined orbits. Can we say anything about inclinations? No hurdle in principle to generalizing the proposed mechanism to inclined orbits. An attempt to work it out with our orbit-averaged treatment of a razor thin disk potential faces an insurmountable singularity \citep{hep80}. Disk height comes in for rescue but then expressions become arbitrary without a specific prescription for vertical disk structure \citep{hah03}. A fix is to use a local approximation in the averages which regularizes expressions \citep{war81} and allows us to explore eccentricity-inclination dynamics for both axisymmetric and eccentric disks. But we went further. 

Starting with the mass and eccentricity distributions in DM1, we computed the full 3D potential and recovered associated spherical harmonics numerically (Kazandjian, Sefilian \& Touma, in preparation). Then we orbit-averaged spherical harmonics (again numerically), and obtained closed form expressions for any given semi-major axis, and to the desired (arbitrary) order in eccentricity and inclination. A brief explanation of the steps involved is provided in Appendix \ref{appendix2}.
 
With the orbit-averaged mean field of the razor thin eccentric disk in hand (Eq. \ref{eqn:numerical_H3D}), we added the secular contribution of the outer planets (Eq. \ref{eqn:planets_H3D}) to study the coupled eccentricity-inclination dynamics of a particle in a perfectly smooth fashion. With the help of these expansions, we could study off-plane dynamics of TNOs which are clustered in the plane, determine stability to small inclinations, as well as the long term evolution of populations of initially clustered and inclined objects. A brief report on global dynamics follows:
\begin{itemize}
\item As evident in Fig. \ref{fig:fig15}, planar phase space structure (including families of equilibria, their stability, and their behavior as a function of semi-major axis) is recovered quite accurately within this generalized formalism; 
\item An involved \textit{linear} stability analysis confirmed that families of stable planar eccentric equilibria (both aligned and anti-aligned with the disk's apsidal line) are further stable to small perturbations in inclination: this is quite encouraging because it suggests that the flock of stationary orbits that were identified in the plane is maintained when subject to small out-of-plane perturbations.
\item Small-amplitude variations in the inclinations, eccentricities, and longitude of apse around stable coplanar equilibria were numerically shown to maintain \textit{near alignment} in the longitude of apse, all the while the argument of apse and longitude of node \textit{circulate}.
\item Moving to large amplitude variations in inclination: Any temptation to inquire about fixed anti-aligned, eccentric and sufficiently inclined orbits is quickly silenced by the realization that both inner quadrupole and eccentric disk induce retrograde nodal precession. While varying with location in and/or inclination to the disk, the reinforced retrograde precession excludes the possibility of apse aligned orbits that further share the same spatial orientation.
\end{itemize}
We could of course proceed to provide a complete classification of orbital dynamics in the combined field of disk and planets. This is a two degree of freedom problem which is amenable to description in terms of Poincar\'e sections at any given semi-major axis. We think that such an exercise is best relegated to a separate purely dynamical treatment. Instead, we opt to follow populations of judiciously chosen particles over the underlying complex phase space, with a view to characterizing the extent to which our setup can reproduce observed metrics. 

\subsection{Populations over Phase Space} \label{sec:pop}

The reference disk, together with the outer planets, sustains two families of stable coplanar equilibria, one aligned ($\Delta\varpi = 0$) and of low eccentricity, the other anti-aligned ($\Delta\varpi = \pi$) and of large eccentricity. 
Anti-aligned equilibria follow the observed trend of eccentricity with semi-major axis (see Fig. \ref{fig:fig3}). 
It is natural to ask what remains of this trend  when vertical heating is included, and when eccentricity-inclination dynamics kicks in. We spoke of linear stability of planar equilibria to slight inclination change. We further reported results of numerical simulations showing that the long term evolution of perturbed planar orbits, shows stability in inclinations for small enough inclination, while maintaining {\it confinement in the longitude  of the apse}. That would suggest that populations of particles initiated around the islands of planar stability would maintain the planar alignment, though not immediately clear for how long in the presence of nonlinearities.

\begin{table}
\begin{center}
\caption{Heliocentric semi major axis ($a_p$) , perihelion distance ($q_p$), inclination ($i_p$), argument of perihelion (${\omega}_p$), longitude of ascending node (${\Omega}_p$), and longitude of perihelion (${\varpi}_p$) of the clustered TNOs with $a_p>250$AU and $q_p>30$AU considered in this study. Data obtained from the Minor Planet Center.
\label{tab:tab3}}
\begin{tabular}{ccccccc}
\tableline
\tableline
TNO & $a_p$ & $q_p$  & $i_p$  & $\omega_p$ & $\Omega_p$  & $\varpi_p$ 
\\
 & (AU) & (AU)  & $({^\circ})$  & $({^\circ})$ & $({^\circ})$  & $({^\circ})$ 
\\
\tableline
$2012 ~ $ VP$_{113}$ & 260.8 & 80.3 & 24.1 & 292.8 & 90.8 & 23.6 \\ 
$2014 ~ $ SR$_{349}$ & 289.0 & 47.6 & 18.0 & 341.4 & 34.8 & 376.2  \\
$2004 ~ $ VN$_{112}$ & 317.7 & 47.3 & 25.6 & 327.1 & 66.0 & 33.1  \\
$2013 ~ $ RF$_{98}$ & 350.0& 36.1 & 29.6 & 311.8 & 67.6 & 379.4  \\ 
$2010 ~ $ GB$_{174}$ & 369.7 & 48.8 & 21.5 & 347.8 & 130.6 & 118.4  \\
$2007 ~ $ TG$_{422}$ & 483.5 & 35.6 & 18.6 & 285.7 & 112.9 & 38.6 \\
Sedna & 499.4 & 76.0 & 11.9 & 311.5 & 144.5 & 96.0  \\
\tableline
\end{tabular}
\end{center}
\end{table}
With this in mind, we explored the dynamics of populations of particles in the combined orbit-averaged gravitational field of DM1 and the outer planets over the age of the Solar System. Particles were initiated around the AI and A-AI of stable planar equilibria (see Fig. \ref{fig:fig4}) at the semi-major axis locations of seven of the clustered objects listed in Table \ref{tab:tab3} \footnote{Performing the study at all positions in the disk, and not only at the seven considered semi-major axes (Table \ref{tab:tab3}),  does not affect the conclusions drawn from the population studies around AI and A-AI.}. Islands of stability were sampled uniformly in eccentricity, with inclinations assigned uniformly in a $10^{\circ}$ range. The argument of pericenter $\omega_p$ and the longitude of ascending node $\Omega_p$ were picked to guarantee uniform $\Delta\varpi$ sampling in the range $180^{\circ} \pm 20^{\circ}$ for the anti-aligned family, and $ \pm 20^{\circ}$ for the aligned family. This way, we end up with 300 particles at each of the seven observed semi-major axis, and follow their orbits over the age of the Solar system. 

We characterize an orbit's  orientation and eccentricity with the Lenz 
\begin{equation}
\textbf{e}_p = e_p \left( \begin{matrix}  \cos w_p \cos\Omega_p - \cos i_p \sin w_p \sin\Omega_p  \\  \cos w_p \sin\Omega_p + \cos i_p \sin w_p \cos\Omega_p   \\ \sin i_p \sin w_p\end{matrix} \right) = e_p \hat{\mathbf{e}}
\label{eqn:Lenz}
\end{equation}
and  specific angular momentum
\begin{equation}
\textbf{h} = \sqrt{G M_{\odot}   a_p (1-e_p^2)} \left( \begin{matrix}  \sin i_p \sin\Omega_p  \\  -\sin i_p \cos\Omega_p   \\ \cos i_p \end{matrix} \right)
\end{equation} 
vectors. The Lenz vector lives in the plane of a particle's orbit and points to its periapse;  the angular momentum vector is perpendicular to the orbital plane, and its dynamics encodes nutation and precession of the orbit. Orbits can have aligned Lenz vectors while being spread out in node and inclination. Orbits can be spread out in Lenz vector while sharing the same orbital plane. In other words, behavior of both vectors is required for a complete characterization of the degree of spatial alignement of a population of orbits, with the following metrics being particularly useful in that regard \citep{mil17}:
\begin{itemize}
\item The departure of Lenz vector orientation from the mean as captured by 
\begin{equation}
S_\varpi(t) = \sum_{i = 1}^{7} \textbf{e}_m(t) \textbf{.} \hat{\mathbf{e}}_i(t),
\end{equation}
where $\left[\hat{\mathbf{e}}_i(t), i=1, \dots,7 \right]$ are unit Lenz vectors and $\textbf{e}_m(t)$ is their mean unit vector. This definition of $S_\varpi(t)$ allows us to quantify the degree to which the Lenz vectors are clustered about their mean. 
Specifically, if the Lenz vectors of the seven objects coincide with their mean at a given time, then $S_\varpi(t) = 7$.
\item A measure of the anti-alignment with respect to the trans-Neptunian disk as given by
\begin{equation}
A_\varpi(t)  = \textbf{e}_m(t)\textbf{.} \hat{\mathbf{e}}_d,
\end{equation} 
where the disk orientation is fixed such that $\hat{\mathbf{e}}_d = (\cos \varpi_d, \sin \varpi_d , 0 )\big|_{\varpi_d = \pi}$ unless otherwise stated.  
A measure of $A_\varpi(t) = -1 ~ (+1)$ corresponds to configurations where the mean Lenz unit vector of the seven bodies is perfectly anti-aligned (aligned) with that of the disk at a given time. 
\item A measure of clustering in $\varpi_p$, $w_p$ and ${\Omega}_p$ separately as provided by the mean over unit vectors; $\textbf{r}_{\varpi_p}$, $\textbf{r}_{w_p}$ and $\textbf{r}_{\Omega_p}$, that circulate with these angles respectively. When the associated vectors are homogeneously distributed on a circle they have zero mean, while perfect alignment results in their mean having unit length.
\end{itemize}
The objects of interest to us listed in Table \ref{tab:tab3} yield a current value of $S_{\varpi} \approx 4.57$  
and $A_{\varpi} \approx -0.77$, assuming
a disk whose apsidal angle is $180^{\circ}$ away from the mean of the clustered inclined bunch.
Moreover, the measures of $\textbf{r}$ for the group of clustered objects are: $\textbf{r}_{\omega_p} = 0.93,  \textbf{r}_{\Omega_p} = 0.81$ and $\textbf{r}_{\varpi_p} = 0.80$, indicating confinement in both $\omega_p$ and $\Omega_p$ as noted by \citet{bat16}.

Based on extensive orbit integrations of our samples, we learned the following:
\begin{itemize}
\item Particles initiated around the AI (Fig. \ref{fig:fig4} [{\bf b}, {\bf c}]) stay tightly bunched while showing small amplitude variations in their inclination, eccentricity and longitude of apse (see Fig. \ref{fig:fig5}). Indeed, we find a time-averaged value of $\textbf{r}_{\varpi_p} = 0.989 \pm 0.011$ indicating strong $\varpi_p$-confinement which is maintained by the opposite circulation of $\omega_p$ and $\Omega_p$ as reflected in $\textbf{r}_{\omega_p} = 0.041\pm0.009 $ and $\textbf{r}_{\Omega_p}= 0.040 \pm 0.003$. In other words, the orbit structure that is expected to self-consistently reproduce the planar disk is stable enough to inclined motion to hold the promise of sustaining a thick version of that disk. In Fig. \ref{fig:fig6}, we show ensemble averaged behavior of $A_{\varpi}(t)$ and $S_{\varpi}(t)$ which supports the conclusion above, with the time-averaged $S_{\varpi} \sim 6.85 $ and $A_{\varpi} \sim 0.98$ indicating Lenz vector confinement, together with disk alignment in the neighborhood of the AI.
%
\begin{figure}
\epsscale{1.2}
\plotone{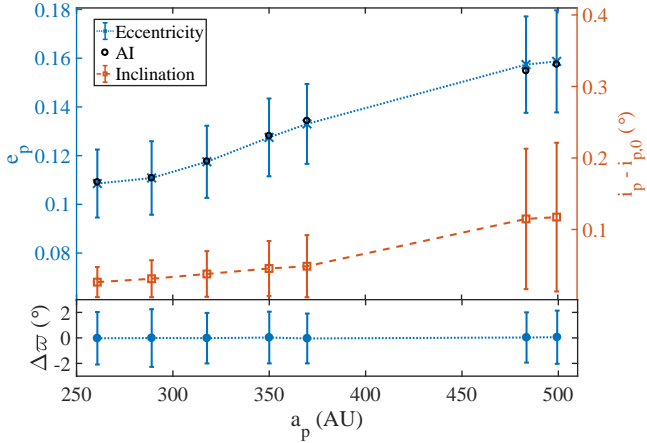}
\caption{The average behavior of eccentricity, inclination and longitude of periapse, and excursions thereabout, for a flock of objects initiated in the neighborhood of the AI. The top panel shows particles at all semi-major axes executing slight excursions in eccentricity and inclinations: values of $e_p$ remain close to the planar equilibria while inclinations oscillate around the initial conditions $i_{p,0}$.
The bottom panel shows how all these particles maintain alignment with the disk with $\Delta\varpi \approx 0^{\circ}$, with negligible spread, over the relevant range in $a_p$. 
\label{fig:fig5}}
\end{figure}
%
\begin{figure}[h!]
\epsscale{1.2}
\plotone{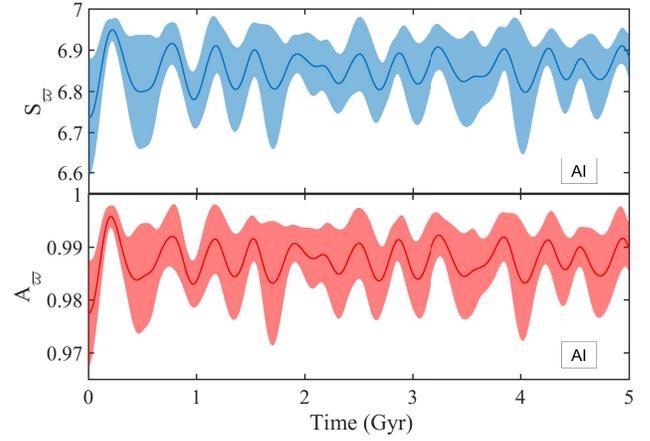}
\caption{
Time evolution of $S_{\varpi}$ and $A_{\varpi}$ for objects initially near the AI.  The calculation is based on an ensemble of 10 sets of particles, each set consisting of 7 particles randomly picked within the AI at each of the considered semi-major axes.  The thick lines represent the ensemble averaged values and the shaded regions enclose the spread around the average. 
\label{fig:fig6}}
\end{figure}
%
\item Populations of particles initiated around the A-AI (Fig. \ref{fig:fig4}{\bf b}) show greater complexity as a function of semi-major axis and inclination.
Particles with semi-major axes $250 \preceq a_p \preceq 350$ AU librate around the planar island with slight excursion in inclination and eccentricity for all initial inclinations (see Fig. \ref{fig:fig7}): linear stability translates into longterm stability in this case, with low inclination orbits maintaining Lenz vector alignment all the while displaying a spread in node and periapse. Indeed, our simulations show that around $93\%$ of all the considered orbits with $a_p \preceq 350 AU$ are strongly confined in $\varpi_p$ with $\textbf{r}_{\varpi_p} > 0.9$ while the node and periapse circulate ($\textbf{r}_{\omega_p} \preceq 0.05$ and $\textbf{r}_{\Omega_p} \preceq 0.03$). 
On the other hand, particles with $a_p \succsim 350$ AU show long term behavior which depends on the initial inclination:
{\bf i.} Particles launched with $i_p \precsim 4- 5^{\circ}$ are stable to off-plane motion, similar to the objects at $a_p \preceq 350 AU$; {\bf ii.}
In contrast, particles launched with $i_p > 5^{\circ}$ show large amplitude variations in $e_p$ and $i_p$, with inclinations growing somewhat erratically to values higher than $20^{\circ}$, all the while  $e_p$ evolves to smaller values. Such particles still show significant clustering in $\varpi_p$ with $\textbf{r}_{\varpi_p} \sim 0.6 $, though weaker than what is observed with stably inclined populations.
\end{itemize}
As evident in Fig. \ref{fig:fig7} and Fig. \ref{fig:fig8}, particles initiated with low inclinations ($i_p < 5^{\circ}$) around the A-AI remain clustered and anti-aligned with the disk, with small amplitude variations in their eccentricities and inclinations. Their stability guarantees the survival of a puffed up version of the backbone of coplanar anti-aligned equilibria  ($\Delta\varpi=\pi$) whose $e_p-a_p$ behavior is consistent with the observed TNOs.
%
%
\begin{figure}
\epsscale{1.2}
\plotone{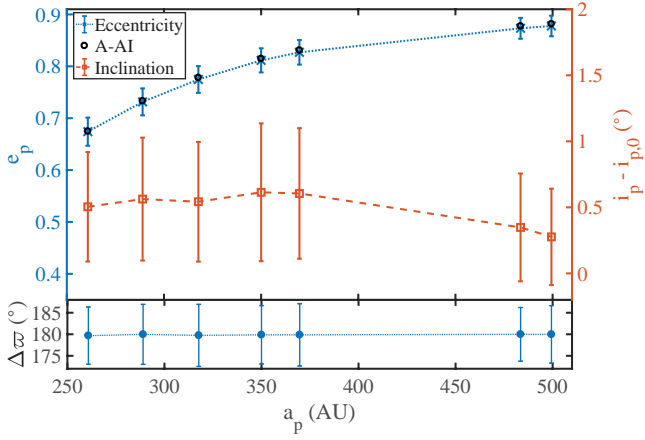}
\caption{Average values of eccentricity, inclination and longitude of periapse, and spread thereabout, for objects initiated near the A-AI with initial inclinations $i_{p,0} < 5^{\circ}$. It is evident that slightly inclined particles initiated around the A-AI exhibit an average behavior of $e_p - a_p$ consistent with the planar equilibrium profile. At the same time, as shown in the bottom panel, the orbits remain clustered in $\varpi$ and anti-aligned with the disk,  with $\Delta\varpi \sim 180^{\circ}$ at all $a_p$.
\label{fig:fig7}}
\end{figure}
%
\begin{figure}[h!]
\epsscale{1.2}
\plotone{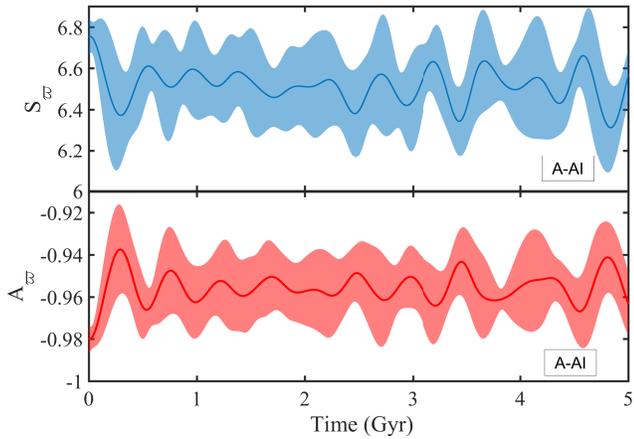}
\caption{Time evolution of the ensemble averaged $S_{\varpi}$ and $A_{\varpi}$ (thick-lines) and the spread thereabout (shaded regions) for objects sampled near the A-AI,  with $i_p(t=0) < 5^{\circ}$. Clustering of Lenz vectors is maintained at all times ($S_{\varpi}(t) \sim 6.5 $) together with disk anti-alignement ($A_{\varpi}(t) \sim -0.96 $). 
\label{fig:fig8}}
\end{figure}
%

\subsection{Clones of Observed TNOs} \label{sec:clones}

We probe the orbital evolution of ``clones" of observed TNOs (Table \ref{tab:tab3}) over the age of the Solar System.
At each of the considered semi-major axes, we build samples of 300 particles with orbital elements randomly picked in the neighborhood of the observed ones ($\delta e = 1\% e_{obs}$ and $\delta  i = \delta \omega = \delta \Omega = 5^{\circ}$). The disk is again coplanar with the outer planets and anti-aligned with the mean apse direction of the clustered TNOs.
%
\begin{figure}
\epsscale{1.2}
\plotone{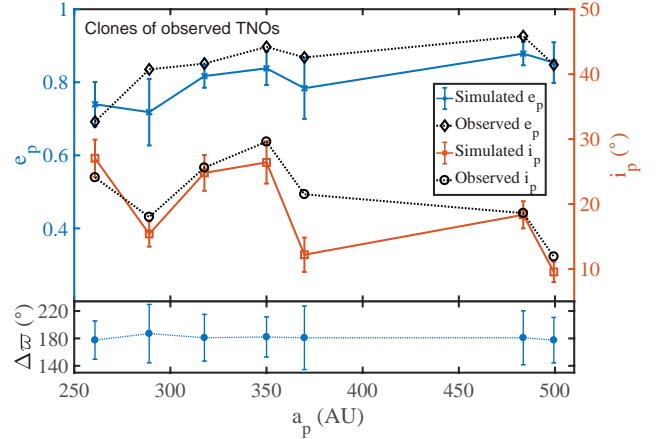}
\caption{The average eccentricity, inclination, and longitude of apse along with their time-averaged spread for all the ``successful clones" in our simulations. The top panel reveals reasonable agreement between observed and average values of both $e_p$ and $i_p$. The bottom panel indicates that clustering and anti-alignment with the proposed disk (DM1) is maintained at all considered values of $a_p$.
\label{fig:fig9}}
\end{figure}
\begin{figure}[h!]
\epsscale{1.2}
\plotone{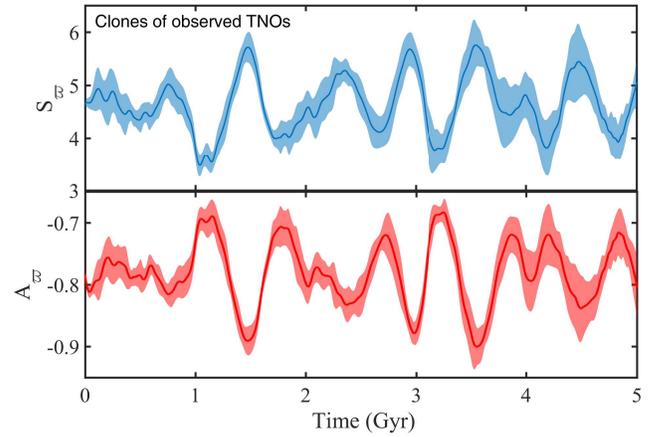}
\caption{Evolution of both $S_{\varpi}$ and $A_{\varpi}$ as a function of time for clones of the observed TNOs.
The calculation is done with ten ensembles of seven particles each, randomly picked from the population of SCs at each of the seven considered values of $a_p$. Thick lines represent the mean, and the shaded regions the spread about that mean.  $S_{\varpi}(t)$ and $A_{\varpi}(t)$ oscillate around their initial values indicating $\varpi_p$-confinement which, on average, is $180^{\circ}$ away from the fixed disk apsidal line.
\label{fig:fig10}}
\end{figure}

We find that more than $60 \%$ of the clones maintain a perihelion distance which is larger than the orbital radius of Neptune at all times. We dub these objects ``successful clones" (SCs for short) and analyze their orbital evolution to conclude that:
\begin{itemize}
\item SCs follow quite closely the eccentricity and inclination of their progenitors (see Fig. \ref{fig:fig9});
\item SCs, on average, maintain anti-alignment with the disk apsidal line, while showing slight oscillations in the longitude of apse around the mean; see Fig. \ref{fig:fig9}. Considering all successful simulations we find $\textbf{r}_{\varpi_p} \approx 0.785$ which compares well with that of the observed bunch; $\textbf{r}_{\varpi_p} = 0.80$; 
\item SCs show no confinement in $\Omega_p$ and $\omega_p$, with $\textbf{r}_{\Omega_p}  = 0.020 \pm 0.006$ and $ \textbf{r}_{\omega_p} = 0.033 \pm 0.019$.
\end{itemize}

Computing $A_{\varpi}(t)$ and $S_{\varpi}(t)$ as before, except with ensembles of SCs, we recover $A_{\varpi} \approx -0.78 \pm 0.03$ and $S_{\varpi} \approx 4.63 \pm 0.34 $ (refer to Fig. \ref{fig:fig10} for the full behaviour of both metrics) . The latter is in agreement with $S_{\varpi}(t=0) \approx 4.57$ for the observed TNOs while the former is consistent with the expected value of $A_{\varpi}(t=0) \approx -0.77$ assuming that the mean apsidal angle of the observed TNOs is $180^{\circ}$ away from that of the hypothesized disk.

In short, the simulations we carried out show that the envisioned disk of trans-Neptunian icy bodies (DM1 to be specific) can provide a fair amount of $\varpi$-confinement for particles whose orbits are seeded in the neighbourhood of the observed clustered TNOs. 

\section{Variations on a Theme} \label{section:4}

Given uncertainties, about disk mass, eccentricity, self-consistent precession,...etc, we thought it
reasonable to explore a range of disk properties around the fiducial ones adopted in what
preceded. Below is a brief account of what we learned, supplemented by the appropriate
figure, when called for. These variations will be assessed with observed properties
and disk self-consistency in mind. They will serve to inform our discussion of how a disk
of the desired properties forms in the first place. 

{\bf Disk Mass.} More massive disks, all else being kept the same, maintain planar equilibria of higher eccentricity, which as is clear by now, will carry over to properties of spatially aligned populations. 
\begin{figure}[h!]
\epsscale{2.3}
\plottwo{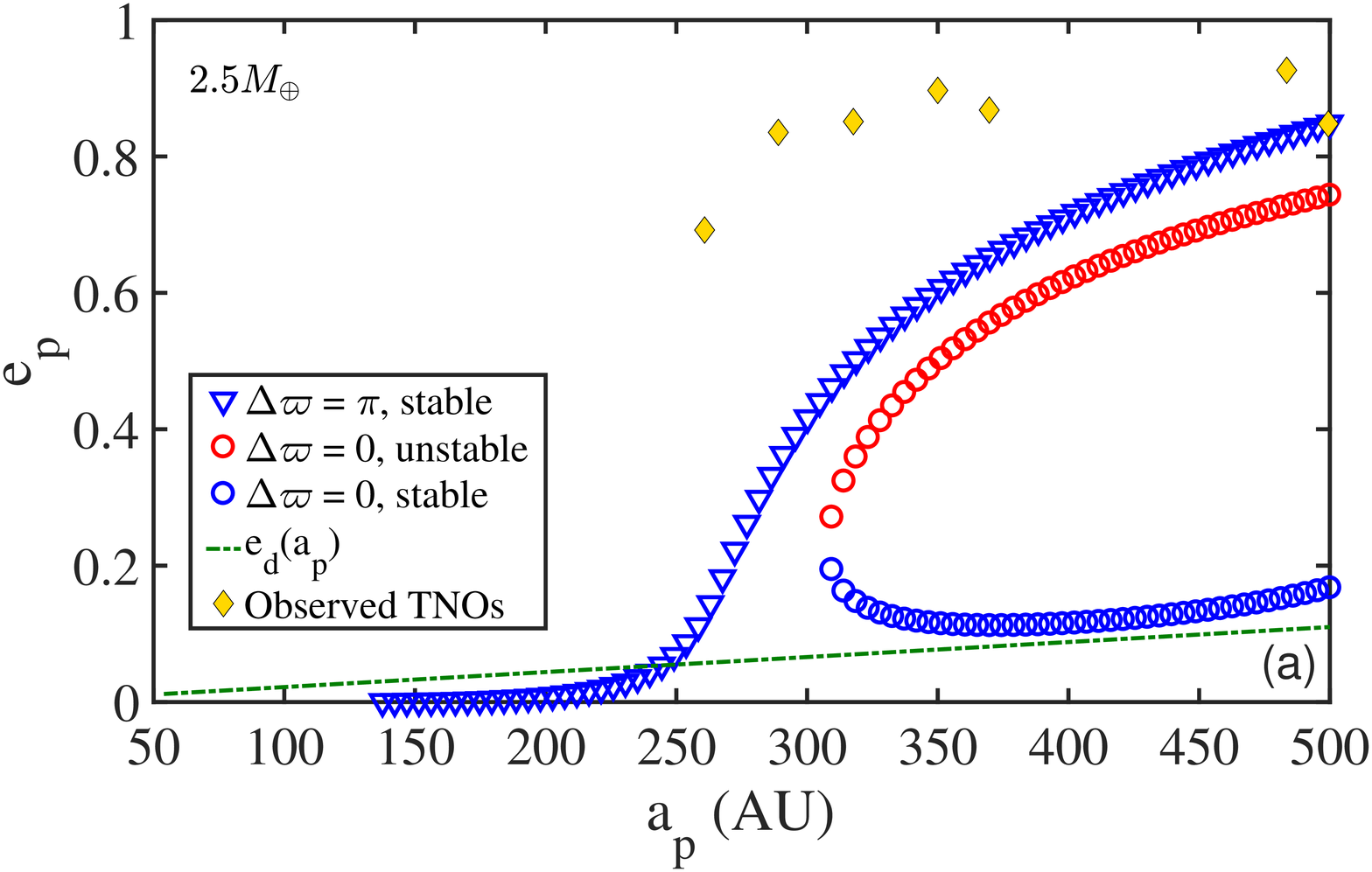}{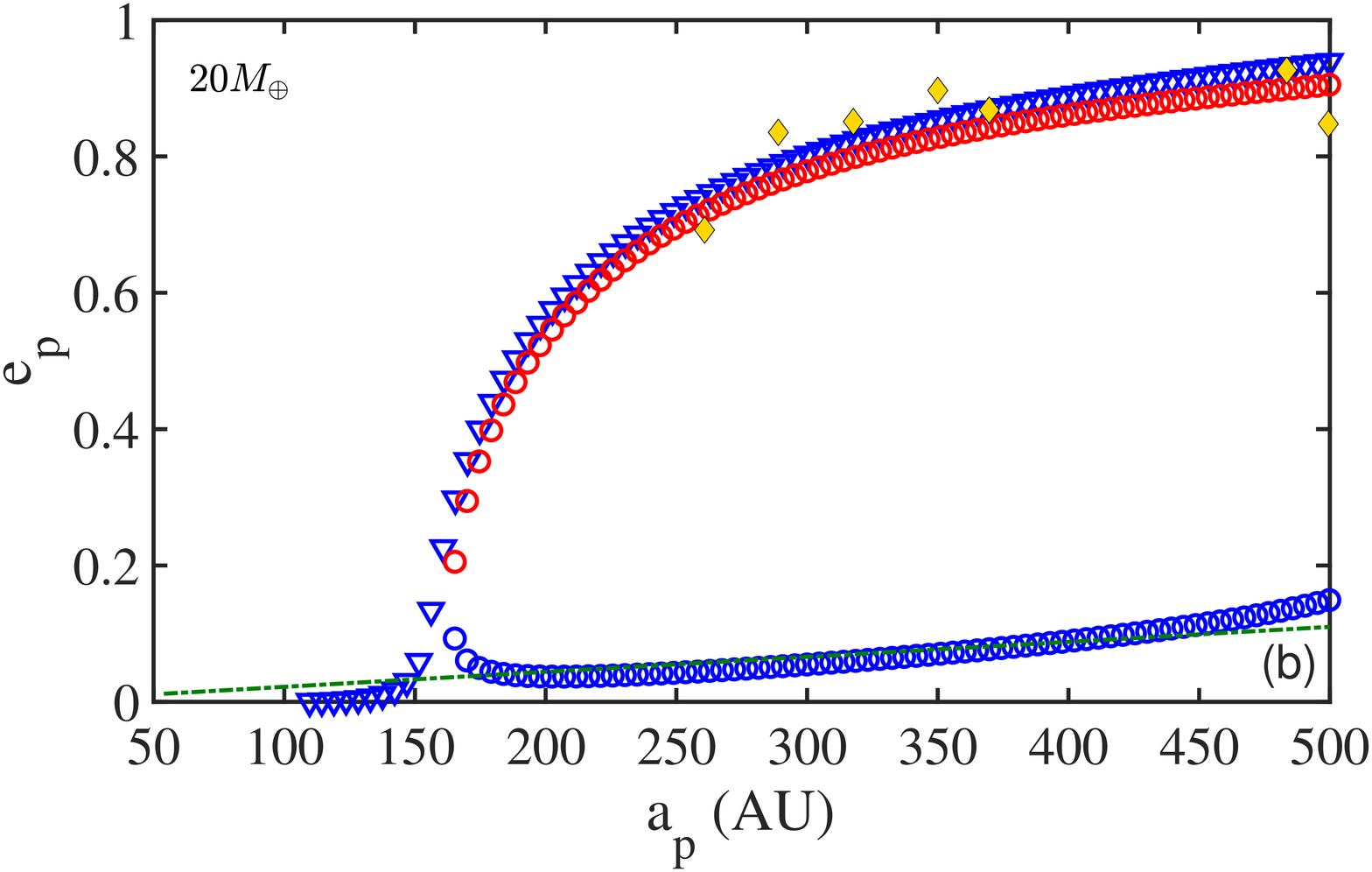}
\caption{Coplanar families of stable and unstable equilibria sustained by disk models DM2 and DM3 (see Table \ref{tab:tab1}). DM2 and DM3 are identical to the fiducial DM1 except that DM2 is less massive with $2.5 M_{\earth}$ (panel {\bf a}) and DM3 more massive with $20 M_{\earth}$ (panel {\bf b}). Evidently, increasing $M_d$ drives up the equilibrium $e_p$ while maintaining the stability of the three families, in agreement with our expectation (see Eq. \ref{eqn:ep-Md_effect}). Furthermore, and as evident in panel {\bf b}, massive disks provide a supply of aligned orbits with $e_p \sim e_d$ over a broader range of $a_p$ .
\label{fig:fig11}}
\end{figure}
Figure \ref{fig:fig11} illustrates the effect in disks that are less and more massive than the adopted reference disk (DM2 and DM3 in Table \ref{tab:tab1}).  This behavior is not too difficult to recover from model equations \ref{eqn:ldot} and \ref{eqn:wdot}, which reduce to
\begin{equation}
e_p \simeq \bigg[ 1- c \,\, \times  M_d^{-\frac{2}{5}} \times a_p^{\frac{2}{5}(p-4)}\bigg]^{0.5}
\label{eqn:ep-Md_effect}
\end{equation}
under the assumption of axisymmetry ($e_d = 0 $) where the constant $c = f(p) \times \big(a_{out}^{2-p}   \sum_{i = 1}^{4 } m_i a_i^2\big)^\frac{2}{5} > 0  $.\footnote{Important to note that Eq. \ref{eqn:ep-Md_effect} captures the trend of growing equilibrium $e_p$ with $a_p$ although the relation was derived for $e_d = 0$. Also note that increasing $a_{out}$ (at constant $M_d$ with $p<2$) tends to lower equilibrium eccentricities.}

Thus, a better fit with the observed eccentricities (ignoring eccentricity-inclination dynamics) can be achieved with a disk mass which is higher
than the one adopted in our analysis (see Fig. \ref{fig:fig11}{\bf b}). Furthermore, the bifurcation of equilibria into aligned and anti-aligned families is seeded earlier in a massive disk, such as DM3, implying that such disks will have a supply of aligned orbits ($\Delta\varpi = 0$) over a broader range of semi-major axes, with which to build themselves up! If explanation be required, the reader should keep in mind that a more massive disk allows for 
stronger precession, hence the ability of low eccentricity orbits to withstand the differential precession induced by the planets, at smaller semi-major axis than otherwise possible with a lighter disk - see Figure \ref{fig:fig11}. 
\begin{figure}[h!]
\epsscale{1}
\plotone{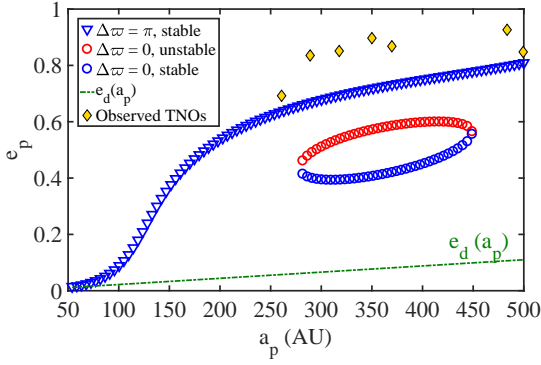}
\caption{The equilibrium TNO families sustained by a disk model analogous to DM1 but with more mass concentrated in the inner parts, $p = 2.5$ (DM4 in Table \ref{tab:tab1}). It is evident that although the assumed disk model gives rise to a coplanar stable family of anti-aligned high $e_p$ orbits, such a disk cannot harbour TNO orbits aligned with the disk such that $e_p \sim e_d(a_p)$.
Such is the case in all disks with mass dropping outwards ($p>2$).
\label{fig:fig12}}
\end{figure}

So far we have taken for granted disk models with mass increasing with $a_p$. We now ask how would things differ in a disk with mass dropping outwards. As evident in Figure \ref{fig:fig12}, in a disk which is otherwise analogous to the fiducial model but with $p = 2.5$ (DM4 in Table \ref{tab:tab1}), it appears impossible to support orbits which are aligned with the disk, and having $e_p \sim e_d(a_p)$. Such (apse-aligned) disks appear unable to sustain themselves, and will not be discussed any further. 

{\bf Disk Eccentricity.} Here one can change both the eccentricity profile (via $q$) as well as the eccentricity at the outer edge of the disk (via $e_0$), for a given profile. 
Reporting on our thorough exploration of the rich set of bifurcations that obtain as a function of disk eccentricity and their implication for the structure of the disk itself will take us too far afield.  Suffice it to say that increasing the outer-edge eccentricity in negative-$q$ disks (i.e. $de_d/da_d > 0$) or adopting eccentricity profiles which drop outwards, keeping all else invariant, introduces greater complexity in the structure of both anti-aligned and aligned planar equilibria, but unfortunately at the cost of loosing those disk-aligned orbits which we believe will be essential in any self-consistent reconstruction of an eccentric disk.
We find that for disks which are so structured that the bulk of their mass is in the outer parts, a disk eccentricity of $e_0 \approx 0.20 - 0.25$ (with \textit{negative} $q$) is the \textit{maximum} that can be tolerated before the eccentricity behavior of the disk-aligned family no longer follows that of the underlying disk. 
\begin{figure}[h!]
\epsscale{2}
\plottwo{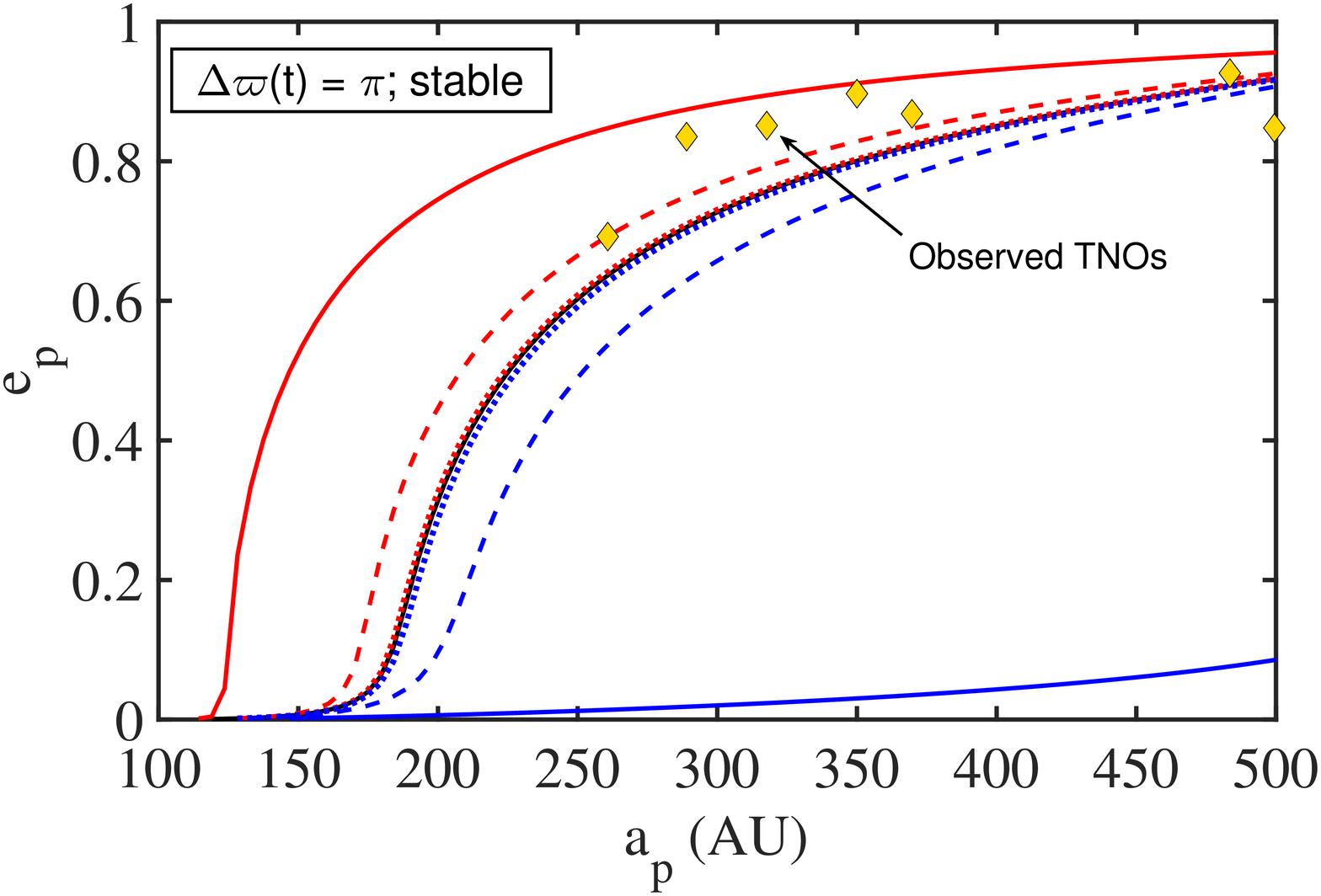}{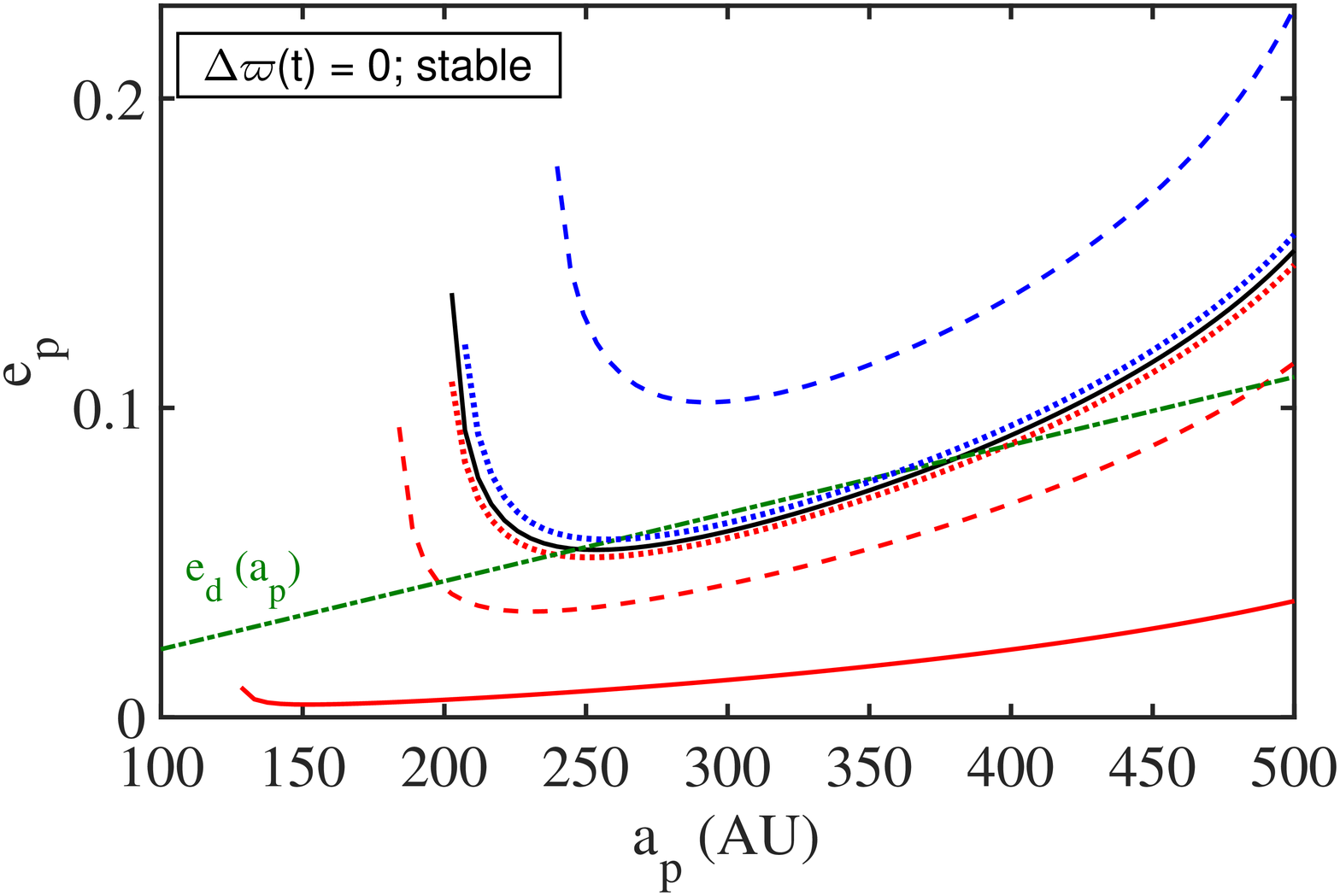}
\caption{The effect of rigid disk precession, both retrograde (in blue) and prograde (in red), 
on the two coplanar stable equilibrium TNO families sustained by non-precessing DM1 (in black).
Precession rates of increasing order, 0.02, 0.2, and $2\times10^{-8} \text{yr}^{-1}$, are depicted by dotted, half-dashed and full lines respectively. The results show that prograde disk precession increases (lowers) the eccentricity of the high(low)-$e_p$ family; one has the opposite effect with $\dot{\varpi}_d < 0$, and this with increasing severity for larger $|\dot{\varpi}_d|$. 
For instance, equilibrium orbits aligned with the disk $(\Delta\varpi(t) = 0)$ are not sustained if $\dot{\varpi}_d = - 2 \times 10^{-8} \text{yr}^{-1}$.
Note that the stability of each equilibrium family is maintained in precessing disks.
\label{fig:fig13}}
\end{figure}

{\bf Disk Precession.} A self-gravitating eccentric disk, which is further torqued by the outer planets, is likely to precess as a whole. The actual rate of precession of a saturated, nonlinear, eccentric mode is difficult to ascertain. The timescale associated with self-sustained precession is on the order of the secular timescale in the disk, $\simeq (M_{\odot}/M_d)\times T_{Kepler}$, and comes out to $\simeq 10^{10}$ years for a circular TNO orbit in a 1 $M_{\earth}$ axisymmetric trans-Neptunian disk\footnote{This characteristic timescale can be reduced by at least an order of magnitude upon accounting for the disc and TNO eccentricities (see Eq. \ref{eqn:wdot}) for a given disc mass. Furthermore, more massive discs drive faster precession since $\dot{\varpi}_p \propto M_d$ (see Eq. \ref{eqn:wdot_approximation}).}.

This is then superposed with differential precession induced by the outer planets, with a timescale of  $10^{10}$ years.
Actually, we can write the contribution of the giant planets to the total TNO precession rate as 
\begin{equation}
\dot{\varpi}_p \big|_{planets} \approx +1.93\times 10^{-10} \text{yr}^{-1} \bigg(  \frac{500 AU}{a_p}  \bigg)^{3.5} 
\end{equation}
for circular TNO orbits.

We explore the structure of equilibria in uniformly precessing disks at three progressively faster pattern speeds (both prograde and retrograde). The results are shown in Figure \ref{fig:fig13}.  For the disk mass being considered, it is evident that with prograde precession, agreement with the observed eccentricity profile improves at the risk of loosing dynamical support from the eccentric aligned and stable family of orbits which acquire lower values of $e_p$ with increasing $\dot{\varpi}_d$.  On the other hand, retrograde precession worsens agreement with the observed family, while shifting the eccentricity profile of the aligned family to eccentricities that are too large to sustain the precessing disk. 

That the desirable features of our fiducial disk are disturbed by imposed precession is of course not surprising. Its properties were optimized under the assumption of zero precession. But now that we have a sense of the effect of uniform precession, we can consider scenarios in which we optimize over disk properties and precession simultaneously.  In particular, one can foresee a lower mass disk undergoing prograde precession while at the same time matching observed high eccentricity orbits, and sustaining a family of stable aligned orbits over a broader range of semi-major axes. 

\section{Discussion} \label{discussion}

Our proposition, with its pros and cons, is perhaps not as singular in the context of planetary system formation as the Planet Nine hypothesis. Still the ingredients
that go into it, the origin of the disk, its mass, its eccentricity, as well as the self-consistent maintenance of the disk itself, require a closer look which we attempt below.

{\bf Disk mass:} There is to be sure much uncertainty concerning the mass that lies beyond Neptune,
let alone question of eccentricity, and self-organization of that mass. We require an eccentric, lopsided, equilibrium disk (precessing or not) of 10 Earth masses or so.
Standard pictures allow for at most a few tenth of Earth masses to be scattered in that region in a primordial disk of planetesimals \citep{gla11, sil18}.
 
Arguments that put a few tenth of Earth masses in that region are either based on extrapolations of observed size distributions, or on numerical simulations of a scattered disk that would invariably allow for such low amounts in that region. As noted in the text, such low masses can contribute to $\varpi$-confinement but they make for eccentricities in disagreement with what is observed (at least in the coplanar, non-precessing disk case - see Fig. \ref{fig:fig11}{\bf a}). But the question is how serious are these constraints? Well the distributions themselves are poorly constrained in their tail, and the dynamical arguments constrained by primordial assumptions which may or may not be legitimate. There are of course alternatives considered in the literature.  \citet{hil81} envisions an intermediate zone between the Kuiper belt and the Oort cloud which is expected to harbor up to a few tens of Earth masses. Then there are suggestions that massive planetesimal disks may be a natural outcome of planet formation processes \citep{ken04, eri18, car17}.

Most important for us however is the exercise of \citet{hog91}
who consider the question of hidden mass, and its gravitational signature
and conclude that by looking at planetary motion and more importantly at
cometary orbits, it is not unreasonable to expect up to a hundred or so Earth
masses in the region in question. The exercise has not been revisited since
(Tremaine, Private communication), though related questions were recently
examined in relation to the Planet Nine hypothesis \citep{fie16,bai16, lai16}.

With the above in mind, it would seem that  a massive trans-Neptunian debris disk is not securely ruled out, hence our suggestion:  rather than lump the perturbing mass in a 10 Earth mass planet, and then find a way to push it out on an inclined and eccentric orbit \citep{ken16, eri18, par17}, allow for the less daring hypothesis of a distribution of coplanar trans-Neptunian objects with a total of $10 M_{\earth}$ and see what it does for you.

{\bf Disk eccentricity: }  Of course, the existence of eccentric particle distributions with inner quadrupolar forcing is not foreign to the Solar System with Uranus's $\epsilon$ ring providing an early example of the type \citep{gol79}. In that context, it was argued that self-gravity provides resistance to differential precession induced by the planet's quadrupole to maintain an eccentric equilibrium configuration for the ring. 
Similar arguments in favor of self-gravitating eccentric distributions are brought to bear on the lopsided double nuclei of galaxies \citep{tre95, pei03}. Such a mechanism may also structure eccentric circumbinary and/or circumprimary protoplanetary disks \citep{paa08, kle08, mes12}. Here, we gave evidence for a family of aligned and moderately eccentric orbits which promises to self-consistently build  the disk that maintains it! So, little that is unusual about an eccentric disk though the details of its origin remain to be explored. 

{\bf Origin:} Observations of ring systems, extra-solar debris disks, stellar disks, as well as theoretical models and associated simulations suggest that eccentric disks are ubiquitous, and are rather easy to stimulate and apparently easy to sustain (with and without inner quadrupolar forcing). There are as many propositions for the origin of self-gravitating eccentric disks as there are dynamicists working in various contexts and at various scales: 
perturbation by passing objects \citep{jac01}; dynamical instabilities that afflict low angular momentum (perhaps counter-rotating distributions) \citep{tou02, tre05, kau18}; and
forcing by eccentric inner or outer binary companion \citep{kle08, paa08, mar09, mes12, pel13}.

In numerical experiments with an eccentric, self-gravitating, narrow ring-like disk, \citet{mad16} noted an inclination instability which was accompanied with a pattern of alignment in argument of periapse. The authors took that as an indication that the process may underly the inclination-eccentricity behavior of the observed clustered TNOs. Intriguing though the proposition maybe, it suffers from various limitations: {\bf a}- simulations do not allow for inner quadrupolar forcing by the planets, or earlier their being embedded in massive gaseous disk; {\bf b}- simulations are not pursued long enough to follow the unfolding of the instability, and its eventual relaxation; {\bf c}- hard to imagine how to form a disk of the required mass in the envisioned hot kinematic state. 

It is likely that inner quadrupolar forcing, if strong enough, can quench the inclination-eccentricity instability altogether, a suggestion which is motivated by related effect in Kozai-Lidov type instability. 

As to the observed pattern of alignment in argument of periapse, it is surely a transient of an in-plane instability, which is expected for high eccentricity disks of the type considered \citep{kau18} and which ultimately relaxes into a lopsided uniformly precessing state of lower mean eccentricity. Gauss wire numerical simulations (Kazandjian, Sefilian \& Touma, in preparation) confirm our expectations, with the disk of \citet{mad16} relaxing into a thick lopsided uniformly-precessing configuration in the presence of outer planet quadrupolar forcing, and remaining axisymmetric when we allow for outer planets which are ten times more massive. 

So while we believe the clustering mechanism of \citet{mad16} is simply a transient which dissolves in time, it seems to do so in just the right sort of self-gravitating, eccentric, thick and uniformly precessing disk that in combination with the outer planets is expected to sustain anti-aligned orbits with behavior comparable to what is observed! The difficulty of course is that the initial conditions for the required instability (bias towards low angular momentum, highly eccentric orbits) seem far from what is expected of distributions of planetesimals at formation. Promising in its ultimate state, but somewhat unlikely in its origin!

{\bf Comment on disk self-consistency:}  Our proposition is predicated on the properties of an idealized disk and its gravitational impact on test-particles that are embedded within it. We showed how a power-law disk can support stable equilibrium  families of eccentric orbits which align with the lopsidedness of the disk, as they reproduce its eccentricity profile. We further showed how, in such a disk, particles which librate in the AI are stable to off-plane perturbations, maintaining disk-alignment. This is all encouraging, in the sense that it suggests that a fully self-consistent thick, lopsided and precessing disk can be constructed.

We would very much like to carry over our dynamical analysis to self-consistent equilibrium disks. For now, we note that in such disks a dispersion of apse directions will surely replace the apse-aligned eccentricity profiles of the present work. Apse dispersion, in the same razor thin disks, will mainly contribute to enhance the potential contribution of the axisymmetric mode over the lopsided one. Such relative adjustments are expected to leave the present qualitative picture pretty much unchanged, all the while inducing variations in the eccentricities of the various equilibrium families. The extreme of course is a disk that is hot enough to have a uniform distribution in the apses, an axisymmetric disk which, depending on its radial density profile, will sustain a degenerate family of equilibria. Slight non-axisymmetry will then break the degeneracy, and nucleate families of aligned and anti-aligned equilibria akin to the ones shown in Fig. \ref{fig:fig3} and Fig. \ref{fig:fig11}. 

{\bf Comment on odd TNOS:}   Currently, three of the eccentric, inclined TNOs with $a_p > 250$ AU and $q_p > 30$ AU fall outside of the $\varpi_p$-confinement which we have sought to account for in terms of an anti-aligned massive eccentric trans-Neptunian disk. Here we briefly review how these objects were analyzed with Planet Nine in the picture, as we further situate them within the phase space structured by DM1 and giant planets:
\begin{itemize}
\item $2013 ~ \text{FT}_{28}$ \citep{she16} and  $2015 ~ \text{KG}_{163}$ \citep{sha17b}: These two objects have apse-orientations which are nearly anti-aligned with the clustered bunch of ten.  For Planet Nine activists, the detection of these two objects was reassuring, for it was understood early on that an eccentric and inclined super-Earth will shelter stable eccentric objects with $\Delta\varpi =0$, i.e. which are planet-aligned in apse \citep{bat16, beu16}.
\citet{she16} take $2013 ~ \text{FT}_{28}$ as symptomatic of a larger cluster (dubbed the ``secondary cluster") of TNOs which are stabilized in aligned orientations by/with Planet Nine.  
\citet{bat17} pointed out that the alignment of $\text{FT}_{28}$ and $\text{KG}_{163}$ might be transient with a relatively short lifetime (100-500 Myrs). 
We have here argued that DM1 shelters a family of stable aligned equilibrium orbits of moderate eccentricity which share the disk's eccentricity profile. While discussing planar phase space dynamics (Section \ref{section:fiducial}), we showed how this family of orbits seeds aligned islands of stability (the so called AIs)  in which particles undergo periodic oscillations in eccentricity and $\varpi$ around the parent orbit. We further pointed out that members of the AI (and of the A-AI) find themselves on orbits which bring them close to the unstable aligned orbit, where they will tend to linger in transient disk-aligned states (see Fig. \ref{fig:fig4}). While it is tempting to suggest that  $2013 ~ \text{FT}_{28}$ and  $2015 ~ \text{KG}_{163}$ are in similar such transient states\footnote{
Simulations of coplanar particles with the semi-major axis of  $2013 ~ \text{FT}_{28}$ ($a_p = 310.1$ AU) show lingering around the unstable aligned equilibrium ($\Delta\varpi=0$) for more than $5$ Gyr when those particles are initiated in the AI or the A-AI (of Fig. \ref{fig:fig4}) on orbits with large enough amplitudes to bring them close to that unstable equilibrium. Naturally, such orbits maintain eccentricities akin to that of the unstable equilibrium ($\sim 0.71$; Fig. \ref{fig:fig3}) which is not so different from that of the inclined TNO $2013 ~ \text{FT}_{28}$ ($\approx 0.86$).}, 
lingering around the family of unstable aligned orbits, only further analysis with variants of DM1 (broad enough to include $2015 ~ \text{KG}_{163}$) can decide that.
\item $2015 ~ \text{GT}_{50}$ \citep{sha17b}: a nonaligned TNO, almost orthogonal to the preferred apse orientation and which, for \citet{sha17b}, is yet another indication that $\varpi$ confinement is due to observational bias (more on bias below). For \citet{bat17},  $2015 ~ \text{GT}_{50}$ is one in a class of eccentric objects which are predominantly controlled by Planet Nine on orbits with circulating longitude of the apse. We again refer to the discussion of planar phase space dynamics in Sec. \ref{section:fiducial} to remind the reader that DM1, together with the outer planets, orchestrates a copious population of highly eccentric apse-circulating orbits, at semi-major axes which keep them safely out Neptune's way.  In particular, clones of $2015 ~ \text{GT}_{50}$ within our model demonstrate circulatory behavior in their $\varpi_p$ while undergoing small-amplitude oscillations in their orbital inclination and maintaining perihelion distance marginally larger than Neptune's orbital radius.
\end{itemize}

{\bf Comment on observational bias:}
\citet{sha17b} scrutenized observational bias in the OSSOS sample and concluded that there is no evidence of clustering in $\omega_p$, $\Omega_p$ and $\varpi_p$ distributions.  A similar conclusion was drawn by \citet{law17}. Indeed, \citet{sha17b} report that although the OSSOS survey is biased towards detecting TNOs with $\varpi_p$ near the region of observed clustering, it was able to detect TNOs across all values of $\varpi$, $2015 ~ \text{GT}_{50}$ included \citep{sha17b}.

On the other hand, \citet{bro17} concluded that although the observed sample is not free of biases, the statistical significance of the signal remains solid. Indeed, \citet{bro17} estimates a rather low probability ($\sim 1.2\%$) for the observed sample (with $a_p > 230$ AU) to be drawn from a uniform population.

Controversy over observational bias may or may not remove the need of a shepherding mechanism responsible for the spatial alignment noted by \citet{bat16}.
Additional TNO discoveries will surely help clarify the matter further.  Here, we would like to build on the $e_p-a_p$ relationship we noted and explored in our secular models (for instance, see Eq. \ref{eqn:ep-Md_effect} and Fig. \ref{fig:fig3}) to propose the following: if it proves that further (seemingly) clustered TNOs maintain the $e_p-a_p$ trend currently correlated with dynamical models, then we take that as a strong observational signature favoring secularly induced clustering in the presence of a massive disk, an outer planet or both. Alternatively, if the $e_p-a_p$ distribution  of such objects reveals significant scatter, above and beyond that implied by the eccentricity-inclination dynamics of our models, then we would take that to weaken the case for dynamical clustering, and weigh more in favor of bias in the observed clustering. 

\section{Conclusion} \label{conclusion}

We probed dynamical behavior stimulated by a relatively massive disk of icy bodies in trans-Neptunian space to flesh out a hunch concerning the interplay between the retrograde apse precession induced by such a disk and prograde precession forced by the outer planets: what if the clustered TNO population inhabits regions of phase space where the two effects cancel? 

Analysis of coplanar dynamics yielded a family of eccentric, clustered, and apse frozen orbits, which showed remarkable agreement with the observed eccentricity-semi major axis distribution.  It further yielded a family of low eccentricity orbits, aligned with the disk, which if properly populated is expected to reproduce the disk that helps sustain them: a self-consistency argument which we require for our disk's mass and eccentricity distributions. 

We then allowed for out-of-plane motion and learned that an eccentric disk promotes linear stability to vertical motion, and gave evidence for the persistence of the planar backbone of stable apse-aligned orbits in inclined dynamics. We further analyzed the orbital evolution of the observed population of spatially aligned, eccentric and inclined bodies, without addressing its origin, and concluded that the envisioned self-gravitating disk maintains what we like to think of as robust observables (i.e. eccentricity, inclination, longitude of apse) over the age of the Solar System. 

We carried out orbital simulations over the age of the Solar System while assuming a fixed planetary configuration, and a stationary disk. We ignored a dissipating gaseous disk, planetary migration and/or the  scattering of objects into the region where our disk resides. We were naturally concerned with the range of behavior sustained in the the ``present" phase space of our hypothesized system. However,  a massive gaseous disk could initially quench an eccentricity-inclination instability in a kinematically hot debris disk, then, with its dissipation, the instability kicks in and allows an initially axisymmetric disk to settle into a thick lopsided configuration which could harbor the apse-aligned orbits that we observe (Kazandjian, Sefilian \& Touma, in preparation). Furthermore, migration of planets, and secular resonances sweeping along with them  might play a role in stirring an extended disk into an eccentric configuration\citep[e.g][and references therein]{hah05}\footnote{To the perspicacious reader who wonders about the evolution of equilibrium families with migrating outer planets, we note that such migration will primarily modify the strength of planetary quadrupolar forcing, thus shifting the location of zero net apse precession in the disk, and moderately perturbing the eccentricities of aligned and anti-aligned equilibria (~$\sim 10-20 \%$ in the course of migration).}.

Our endeavor takes observational ``evidence" for granted. \citet{sha17a} cast doubt on the significance of the signal, further arguing that the mere observation of clustered TNOs, when taken at face value, requires a massive  ($\sim 6-24 M_{\earth}$) extended reservoir of TNOs.  We are of course happy to hear about indications for a massive population of TNOs, while our colleagues see it as problematic given the currently favored estimates for mass in this region of the Solar System. These estimates put the total mass at  $\sim 0.1 M_{\earth}$ \citep{gla11}. They are largely based on empirically constrained size distributions with significant uncertainty in their tails. 
We question those estimates as we point to recent global simulations of protoplanetary disks suggesting the production of rather massive ($\succsim 60~M_{\earth}$) planetesimal disks beyond 100 AU \citep{car17}.

The last thorough attempt at dynamical modelling of baryonic dark matter in the outer Solar System was undertaken in the early nineties \citep{hog91}. By considering variations in cometary orbital elements, \citet{hog91} argued for a few (perhaps hundreds) of Earth masses on scales of 100 AU. We like to think that this early exercise [which incidentally was undertaken to carefully examine the evidence (or lack thereof) for a tenth planet] is being revisited piecemeal with Planet Nine in mind \citep{bat16, fie16, bai16}!  We propose that similar such indirect measures be undertaken with an extended moderately eccentric few Earth mass disk perhaps replacing, perhaps combined with, a trans-Neptunian planet.

Of course we can draw comfort in our hypothesis from observations of extra-solar debris disks particularly massive ones around planet-hosting stars\footnote{For instance, the analog of the Kuiper belt around $\tau$ Ceti has been estimated to contain $1.2 M_E$ in $r<10km$ objects \citep{gre04} with four candidate planets orbiting the star in the inner region \citep{fen17}.}, as we hope for the mechanism of $\varpi$-confinement identified in this study to shed light on the dynamics and structure of these disks. 

Ultimately though, we do not have secure and direct observational evidence for our proposed disk, pretty much like we do not have full proof arguments against Planet Nine. Still, we hope to have given sufficiently many dynamical indicators for the game changing role of such a disk in shepherding eccentric TNOs over a broad range of semi-major axes. Of course, a massive eccentric disk could operate simultaneously with a post-Neptunian planet to assure full secular spatial confinement if and when called for.  And the converse is also true in the sense that, with the proper mass distribution and orbital architecture, a disk-planet combination may prove capable of stabilizing orbits in configurations which are difficult to maintain with Planet Nine acting alone.  TNO 2013 $\text{SY}_{99}$ \citep{ban17} provides a case in point. A newly discovered object, its orbit is highly eccentric, apparently clustered with the wild bunch, but unlike its companions nearly in the ecliptic with $\sim 4^{\circ}$ orbital inclination. This object is so tenuously held on its orbit that it is exposed to the randomizing influence of Neptune promoting diffusion at the inner edge of the Oort cloud. Interestingly enough, a Planet Nine like influence is not of much stabilizing help here. In fact, when allowance is made for an inclined eccentric ninth planet, a la \citet{bat16}, all hell breaks loose in the orbital evolution of this curious TNO \citep{ban17}. Well, forgetting about Planet Nine for the moment, and entertaining, as we like to do,  the possibility of a massive trans-Neptunian disk, we naturally find stable anti-aligned coplanar equilibria at a few hundreds AUs, and with nearly the observed eccentricity! In fact, following the orbital evolution of 2013 $\text{SY}_{99}$ under the action of our hypothesized disk, we learned that its current orbit can be sustained over the age of the Solar System,  executing only mild oscillations in inclination and perihelion distance, while maintaining near-alignment in $\varpi$ as the pericenter and node circulate. These two limits, along with the eccentricity-semi major axis distribution which we highlighted, speak in favor of the combined action of a self-gravitating trans-Neptunian disk, together with a trans-Neptunian terrestrial core (e.g. something akin to what was recently suggested by \citet{vol17}, or perhaps the result of a scattering event a la \citet{sil18}). We end with the hope for this combined action to be the subject of parametric studies akin to the ones undertaken with Planet Nine acting alone.

\acknowledgments

This work grew out of a graduate seminar on the Oort Cloud conducted in Fall 2016 at the American University of Beirut. 
Discussions with participants M. Khaldieh and R. Badr are gratefully acknowledged.   The authors would like to thank S. Tremaine, S. Sridhar, L. Klushin and M. Bannister  for helpful discussions,  and an anonymous referee for constructive comments.
We thank M. Kazandjian for making his modal analysis toolbox available to us. 
J.T. acknowledges insightful comments by R. Touma concerning the potential interplay between Planet Nine and a massive eccentric disk.
A.S. acknowledges a scholarship by the Gates Cambridge Trust (OPP1144). Open Access for this article was funded by Bill \& Melinda Gates Foundation.

\appendix
\section{The Disturbing Potential Due To The Disk} \label{appendix1}

We present explicit expressions for the expansion of the orbit-averaged disturbing function (per unit mass) $R_d$ due to a disk composed of coplanar, apse-aligned, and confocal ellipses given in equation (\ref{eqn:Rd}).
The expansion is carried out to fourth order in test-particle eccentricity, generalizing the work of \citet{sil14}, and is valid for arbitrary semi-major axes by the aid of the classical (unsoftened) Laplace coefficients.
The mathematical details and techniques involved; from expanding the disturbing function in eccentricities and retaining the secular part, are discussed in \citet{sef17}.

We consider a finite razor-thin disk, orbiting a central massive object, composed of confocal eccentric apse-aligned streamlines. The non-axisymmetric surface density of such a disk can be written as \citep{sta99},
\begin{equation}
\Sigma(a_d,\phi_d) = \Sigma_d(a_d) \frac{1-e_d(a_d)^2 - a_d e_d^{'} [1 +e_d(a_d)]}{1-e_d(a_d)^2-a_d e_d^{'}[e_d(a_d) + \cos E_d]}
\label{eqn:A1}
\end{equation}
where $E_d$ ($\phi_d$) is the eccentric (true) anomaly of the disk element and $e_d^{'} \equiv \frac{d}{da_d}e_d(a_d)$.
\\
In what follows, we have assumed power-law relations for $\Sigma_d(a_d)$ and $e_d(a_d)$ as given in equations (\ref{eqn:Sigma_d}) and (\ref{eqn:e_d}), although it is possible to use the same methods involved to recover expansions for arbitrary profiles as well as for cases with $\varpi_d = \varpi_d(a_d)$ \citep{sta01,ogi01} .
\\
Our aim is to compute the secular potential $\Phi_d$ experienced by a test-particle embedded in the disk,
\begin{equation}
\Phi_d = - R_d = - G \bigg< \int \frac{\Sigma(r_d,\phi_d) r_d dr_d d{\phi_d}}{\Delta} \bigg>
\label{eqn:R__d}
\end{equation} 
where the integration is performed over the area of the eccentric disk, $<..>$ represents time-averaging over the test-particle orbit, G is the gravitational constant, and 
$
\Delta^2 = r_p^2 + r_d^2 - 2 r_p r_d \cos\theta$
with $r_p$ and $r_d$ being the instantaneous position vectors of the test-particle and disk element respectively such that $(\textbf{r}_p, \textbf{r}_d) = \theta$.
\\
Following the classical formulation developed by \citet{hep80} who first computed $\Phi_d$ for axisymmetric disks without softening the kernel $\Delta$ \citep[see][]{war81}, we make use of the techniques laid down in \citet{sil14} and extend their derivation to fourth-order in eccentricities rendering our forumalae applicable in general astrophysical setups which need not be cases where the assumption of $e<<1$ holds.
\\
After a laborious task of algebraic manipulations (expanding, averaging...) and ignoring constant terms which have no dynamical effect at the secular level, we find that $\Phi_d$ takes the following form
\begin{equation}
\Phi_d = -R_d = - K \bigg[ \psi_1 e_p \cos(\varpi_p-\varpi_d) 
+ \big(\psi_2 + \psi_3 \cos (2\varpi_p-2\varpi_d) \big) e_p^2 + \psi_4 e_p^3 \cos (\varpi_p-\varpi_d) +\psi_5 e_p^4 \bigg]
\label{eqn:FINALformofR}
\end{equation}
where $K = \pi G \Sigma_0 a_{out}^p a_p^{1-p} > 0$ and the dimensionless coefficients $\psi_i$ depend on $\alpha_1 \equiv a_{in}/a_p$, $\alpha_2 \equiv a_p/a_{out}$, and the power-law indices p and q.
To compactify the expressions of $\psi_i$, we use the definition of Laplace coefficients
\begin{equation}
b_{s}^{m}(\alpha) = \frac{2}{\pi} \int\limits_{0}^{\pi} \frac{\cos(m\theta)}{(1+\alpha^2-2\alpha\cos\theta)^{s/2}} d\theta 
\end{equation}
and define the following auxiliary functions
\begin{equation}
I(x,y,z) = \int\limits_{x}^{1} \alpha^y b_1^{z}(\alpha) d\alpha 
~ ~ ~ \text{and} ~~~
D_i^m [f] = \frac{d^m}{d\alpha_i^m} f(\alpha_i) ~\text{for i = 1,2 and m} \succeq 0
\end{equation} 
to write
\begin{eqnarray}
J_1 &=& I(\alpha_1, 1-p, 0 )  + I(\alpha_2, p-2, 0 ) \\
J_2 &=& I(\alpha_1, 1-p-q, 0) + I(\alpha_2, p+q-2, 0)\\
J_3 &=& I(\alpha_1, 1-p-q, 1) + I(\alpha_2, p+q-2, 1)
\end{eqnarray}
These definitions allow us to cast the coefficients $\psi_i$ in the following form:
\begin{eqnarray}
\psi_1  &=& \frac{e_d(a_p)}{2}\bigg\{ -(p+q)(p+q-3) J_3  + \alpha_2^{p+q-1} \big[(2-p-q) D_2^0 b_1^{1} + \alpha_2 D_2^1 b_1^{1} \big] + \alpha_1^{2-p-q} \big[ (p+q-1) D_1^0 b_1^1 + \alpha_1 D_1^1 b_1^1 \big] \bigg\} 
\nonumber 
\\
& +&  e_d(a_p)^2(p+2q)\bigg[ \alpha_2^{p+2q-1}\bigg( D_2^0 b_1^1 + \frac{\alpha_2}{2}  D_2^1 b_1^1   \bigg)  - \frac{\alpha_1^{2-p-2q}}{2}  \bigg(  D_1^0 b_1^1 - \alpha_1 D_1^1 b_1^1  \bigg)	\bigg]         
\label{eqn:psi1}
\\
\psi_2  &=&   \frac{(1-p)(2-p)}{4} J_1     +  \frac{1-p}{2} \alpha_1 D_1^0[\alpha_1^{1-p} b_1^0]   
 -\frac{\alpha_1^2}{4} D_1^1[\alpha_1^{1-p} b_1^0]  + \frac{p-2}{2}\alpha_2 D_2^0[\alpha^{p-2}  b_1^0]    -\frac{\alpha_2^2}{4} D_2^1[\alpha^{p-2}  b_1^0]
 \nonumber \\
 & +&  \frac{e_d(a_p)}{4}\bigg[q(1-p-q)(2-p-q)  J_2     + 2 q (1-p-q) \alpha_1 D_1^0[\alpha^{1-p-q} b_1^0] - q \alpha_1^2 D_1^1[\alpha^{1-p-q} b_1^0]
 \nonumber\\
& + &  2q(p+q-2) \alpha_2 D_2^0[\alpha^{p+q-2} b_1^0]   -q\alpha_2^2    D_2^1[\alpha^{p+q-2} b_1^0]   
+ 2\alpha_2^{p+q}   \bigg(  D_2^1 b_1^0   +\frac{\alpha_2}{2}  D_2^2b_1^0           \bigg)      
 -2\alpha_1^{3-p-q} \bigg( D_1^1 b_1^0     +\frac{\alpha_1}{2}  D_1^2 b_1^0   \bigg)     \bigg]  
 \nonumber \\
& +&  e_d(a_p)^2 \frac{p+3q}{4}\bigg[     \alpha_2^{p+2q}      \bigg(  D_2^1 b_1^0 + \frac{\alpha_2}{2} D_2^2 b_1^0   \bigg)  - \alpha_1^{3-p-2q}
\bigg(  D_1^1 b_1^0 + \frac{\alpha_1}{2} D_1^2 b_1^0   \bigg)            \bigg] 
\label{eqn:psi2}
\\
\psi_3  &=&   e_d(a_p)^2\frac{p+q}{8}   \bigg[3\alpha_2^{p+2q-1} \bigg( D_2^0 b_1^2 + \alpha_2 D_2^1 b_1^2 + \frac{\alpha_2^2}{6} D_2^2 b_1^2  \bigg)  
-\alpha_1^{2-p-2q}\bigg( D_1^0 b_1^2 -\alpha_1 D_1^1 b_1^2 + \frac{\alpha_1^2}{2} D_1^2 b_1^2      \bigg) \bigg]  
\label{eqn:psi3} 
\\
 \psi_4  &=&   \frac{e_d(a_p)}{16}  \bigg[(p+q)^2(1-p-q)(p+q-3) J_3   -(p+q)\big[(p+q-2)(3p+3q-5)-2\big] \alpha_2 D_2^0 [\alpha^{p+q-2}b_1^1] \nonumber \\
 &+& (p+q)(3p+3q-7)\alpha_2^2 D_2^1[\alpha^{p+q-2}b_1^1] - (p+q) \alpha_2^3 D_2^2[\alpha^{p+q-2}b_1^1] + (p+q)^2(3p+3q-5) \alpha_1 D_1^0[\alpha^{1-p-q}b_1^1]
 \nonumber \\
 & +&  (p+q)(3p+3q-1) \alpha_1^2 D_1^1[\alpha^{1-p-q}b_1^1 ]  +     (p+q)\alpha_1^3   D_1^2[\alpha^{1-p-q}b_1^1] + \alpha_2^{p+q+1} \bigg(4  D_2^2b_1^1 +\alpha_2  D_2^3 b_1^1  \bigg)
 \nonumber \\
 &- &  2\alpha_1^{2-p-q}\bigg(  D_1^0 b_1^1 - \alpha_1 D_1^1 b_1^1 - \frac{5\alpha_1^2}{2} D_1^2b_1^1 - \frac{\alpha_1^3}{2} D_1^3 b_1^1  \bigg) \bigg]
 \label{eqn:psi4}
 \\
\text{and}
\\
\psi_5  &=& \frac{p(1-p^2)(2-p)}{64}   J_1 
 -\frac{\alpha_1}{16}p(p^2-1) D_1^0[\alpha^{1-p} b_1^0] - \frac{3\alpha_1^2}{32}(p^2+p) D_1^1 [\alpha^{1-p} b_1^0]  -\frac{\alpha_1^3}{16}(p+1) D_1^2[\alpha^{1-p} b_1^0] - \frac{\alpha_1^4}{64} D_1^3[\alpha^{1-p} b_1^0] 
 \nonumber\\
&+ &   \frac{\alpha_2}{16}(p^3-3p^2+2p) D_2^0[\alpha^{p-2} b_1^0] -\frac{3\alpha_2^2}{32} (p-1) (p-2) D_2^1[\alpha^{p-2} b_1^0]    - \frac{\alpha_2^3}{16}(2-p) D_2^2[\alpha^{p-2} b_1^0]
 -\frac{\alpha_2^4}{64} D_2^3[\alpha^{p-2} b_1^0]     
 \label{eqn:psi5} 
\end{eqnarray}
We note that the disk potential $\Phi_d$ presented here reduces to that of \citet{sil14} upon limiting it to second order in eccentricities and naturally, to that of \citet{hep80} for axisymmetric disks.
\\
We end with a brief remark about the behavior of the coefficients $\psi_i$. The expressions of $\psi_i$ depend on the disk boundaries through $\alpha_1$ and $\alpha_2$ such that the magnitudes of $\psi_i$ diverge when the test-particle is situated nearby the disk edges. However, in the opposite limit, when the test-particle semi-major axis is well separated from the disk boundaries $a_{in}$ and $a_{out}$, we can ignore the edge effects provided that the disk spans several order of magnitude in radius. In such a case ($\alpha_1, \alpha_2 \rightarrow 0$) we can get relatively simple closed form expressions for $\psi_i$ valid for $a_{in} << a_p << a_{out}$ using the series expansions of complete elliptic integrals \citep{gra94} such that;
\begin{eqnarray}
\psi_1 &=& e_d(a_p) \bigg[ \frac{3}{2} - (p+q)(p+q-3) \sum\limits_{n=2}^{\infty} \frac{2n A_n(4n-1) }{(2n-1)(2n+1-p-q)(2n +p+q-2)}    \bigg] 
\label{eq:psi1_expanded}
\\
\psi_2 &=& -\frac{1}{2} + \frac{(1-p)(2-p)}{2} \sum\limits_{n=1}^{\infty} \frac{(4n+1)A_n}{(2n+2-p)(2n+p-1)} \nonumber 
\label{eq:psi2_expanded}
\\
&+& \frac{q e_d(a_p)}{2} \bigg[   1 + (1-p-q)(2-p-q)  \sum\limits_{n=1}^{\infty}  \frac{(4n+1)A_n}{(2n+2-p-q)(2n+p+q-1)} \bigg]
\\
\psi_3 &=& 0 
\\
\psi_4 &=& \frac{ (p+q) (p+q-1) }{8} \psi_1 
\\
\psi_5 &=& \frac{p(p+1)}{16} \psi_2\big|_{q=0}
\label{eq:psi5_expanded}
\end{eqnarray}
where  $ \sqrt{A_n} = \frac{(2n)!}{ 2^{2n} (n!)^2}$.\\

\section{Numerical 3D Potential of a Disk} \label{appendix2}

Here, we present a brief recipe to recover numerically the full 3D gravitational potential generated by a disk formed of coplanar, apse-aligned, eccentric rings. The numerical tool that we developed is indeed general and can be employed to recover the potential due to any configuration of rings (warped, inclined, spiral-armed disks, etc...). The ramifications of this tool will be explored in the future (Kazandjian, Sefilian \& Touma, in preparation).

For the purposes of this study, we distributed N (=1024) coplanar, apse-aligned rings over the range of the disk with specified mass and eccentricity distributions (Eq. \ref{eqn:Sigma_d} and \ref{eqn:e_d}). We then computed the full 3D potential experienced by test-particles by recovering the associated spherical harmonics $Y_l^m(\theta,\phi)$ numerically. Identifying the most dominant modes $a_{l,m}(r)$, we then orbit averaged numerically the arising terms and obtained closed form expressions for the modes in question for any given semi-major axis.

This numerical procedure allows us to express the secular potential of any disk as 
\begin{equation}
\bar{\Phi}_d = \sum_{l, m}\big< a_{l,m}(r) Y_l^m(\theta, \phi) \big> 
\label{eqn:appb_1}
\end{equation}
where $<..>$ stands for time-averaging, and the angles $\theta$ and $\phi$ can be expressed in terms of the usual orbital elements using
\begin{subequations}
\begin{align}
& x = r \sin\theta \cos\phi 
= r [ \cos(\Omega-\varpi_d) \cos(w+f) - \sin(\Omega-\varpi_d) \sin(w+f) \cos(i)    ]     \\
& y = r \sin\theta \sin\phi = r [ \sin(\Omega-\varpi_d) \cos(w+f) + \cos(\Omega-\varpi_d) \sin(w+f) \cos(i)    ]      \\
& z = r \cos\theta = r \sin(w+f) \sin(i)
\end{align}
\end{subequations}
where $f$ is the true anomaly.
\\
Equipped with the relevant dominant modes $a_{l,m}(r)$, we find that the potential due to the reference disk (DM1; Table \ref{tab:tab1}) takes the following form
\begin{equation}
\bar{\Phi}_d =
-\sqrt{\frac{1}{4\pi}}  < a_{0,0}(r) >  
+  \sqrt{\frac{5}{4\pi}}  <a_{2,0} P_2^0(\cos\theta) >  
- \sqrt{\frac{3}{8\pi}} < a_{1,1}(r)  P_1^1(\cos\theta) \cos\phi >  
+ \sqrt{\frac{7}{48\pi}} < a_{3,1}(r)  P_3^1(\cos\theta) \cos\phi >  
\end{equation}
where we have assumed that we have averaged over the orbit of the test particles ( $<..>$ ) and $P_l^m(x)$ are the Legendre polynomials. 
\begin{figure*}[ht!]
\epsscale{0.85}
\plotone{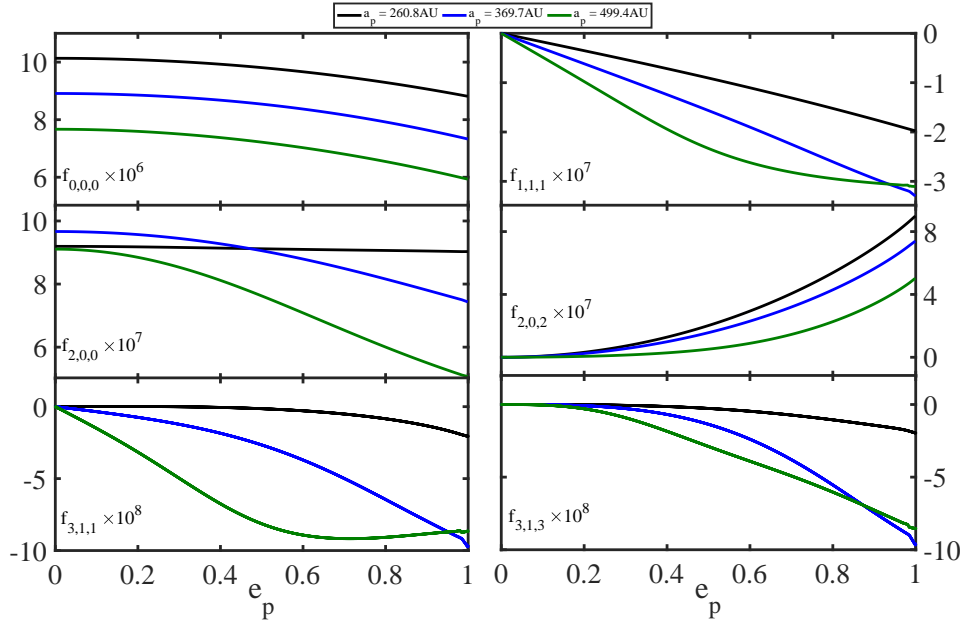}
\caption{The behavior of numerically-averaged functions $f_{l,m,n} = < a_{l,m}(r) \cos (n f)>$ appearing in the disk potential (Eq. \ref{eqn:numerical_H3D}) as a function of eccentricity $e_p$ for orbits with different semi-major axes $a_p$ as shown. The calculation is performed for the fiducial disk configuration DM1 (see Table \ref{tab:tab1}). The chosen values of $a_p$ correspond to semi-major axes at which TNOs are observed (see Table \ref{tab:tab3}). Note that the transition in behavior of $f_{l,m,n}$ with varying semi-major axis is smooth.
\label{fig:fig14}}
\end{figure*}
\begin{figure}[h!]
\epsscale{0.7}
\plotone{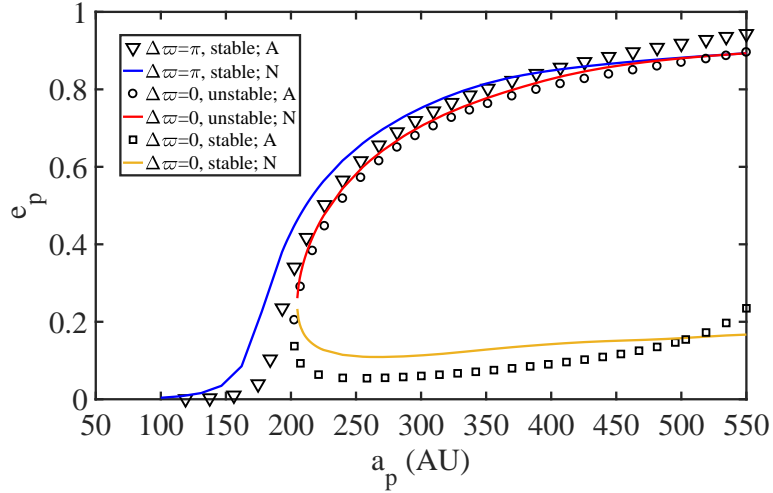}
\caption{Comparison between the coplanar equilibrium families obtained via orbit-averaged disk potential (\textbf{A}; Eq. \ref{eqn:Rd}) and orbit-averaged harmonics (\textbf{N}; Eq. \ref{eqn:numerical_H3D} with $i_p = 0$) of the potential generated by DM1 (Table \ref{tab:tab1}). It is evident that both treatments of the disk potential, when combined with that of the giant planets, yield very similar equilibrium structure, including stability and behavior as a function of semi-major axis.
\label{fig:fig15}}
\end{figure}
After a long, tedious, but straightforward algebra, we express $\bar{\Phi}_d$ in terms of the orbital elements and write 
\begin{eqnarray}
\bar{\Phi}_d &=& F(e_p) + G(e_p)\big[ \cos(\Omega_p-\varpi_d)\cos(w_p) - \cos(i_p) \sin(\Omega_p - \varpi_d) \sin(w_p)   \big] \nonumber 
\\
&+& \frac{3}{4} \sqrt{\frac{5}{4\pi}} \sin^2(i_p) \bigg[   f_{2,0,0} - \cos(2w_p) f_{2,0,2}    +  \sqrt{\frac{35}{48}} \cos(\Omega_p -\varpi_d) \big[ \cos(3w_p) f_{3,1,3} - \cos(w_p) f_{3,1,1}   \big]                   \bigg] \nonumber
\\
&+& \frac{15}{8} \sqrt{\frac{7}{48\pi}} \cos(i_p) \sin^2(i_p) \sin(\Omega_p-\varpi_d) \sin(w_p) \big[3 f_{3,1,1} -  f_{3,1,1}   -2 f_{3,1,3}  \cos(2w_p)    \big]
\label{eqn:numerical_H3D}
\end{eqnarray} 
where $f_{l,m,n} (e_p; a_p) = < a_{l,m}(r) \cos(nf) > $ are computed numerically at a given semi-major axis to the desired (arbitrary) order in eccentricity, and we have defined
\begin{subequations}
\begin{align}
& F =    - \sqrt{\frac{1}{4\pi}} f_{0,0,0} - \frac{1}{2}  \sqrt{\frac{5}{4\pi} }  f_{2,0,0}                 \\
& G = \sqrt{\frac{3}{8\pi}} f_{1,1,1} + \frac{3}{2} \sqrt{\frac{7}{48\pi}} f_{3,1,1}
\end{align}
\end{subequations}
\\
It is noteworthy that all coplanar, apse-aligned, power-law disks share this general form of the Hamiltonian with the only difference being in the functions $a_{l,m}(r)$ which depend on the characteristics of the disk: mass distribution, eccentricity profile and the boundaries.
For reference, in Fig. \ref{fig:fig14} we show the time-averaged behavior of the functions $f_{l,m,n}$ for DM1 as  a function of test-particle eccentricity at three values of semi-major axes.
\\
With the orbit-averaged mean field of the razor thin eccentric disk in hand, we add the secular contribution of the outer planets and write the total Hamiltonian H as 
\begin{equation}
H = \bar{\Phi}_d  - \frac{\Gamma}{6l_p^3} [ 3\cos(i_p)^2 - 1 ]
\label{eqn:planets_H3D}
\end{equation}
where $\Gamma$ is given by equation \ref{eqn:GammaHp}.

This allows us to express the equations of motion governing the dynamics of a TNO under the effect of the considered disk and the giant planets as: 
\begin{eqnarray}
 L_p \dot{\Omega}_p &=&  
- C_i l_p^{-1} \bigg( \frac{\Gamma}{l_p^3}  + \frac{3}{4} \sqrt{\frac{5}{\pi}} T_1   
+ \frac{5}{16} \sqrt{\frac{21}{\pi}} C_h T_2         \bigg)
- S_h S_w l_p^{-1} \bigg[ \frac{15}{32} \sqrt{\frac{21}{\pi}} C_i^2 T_3 
+   \bigg( G -   \frac{5}{32}\sqrt{\frac{21}{\pi}} T_3                      \bigg) \bigg]
\\
L_p \dot{l}_p &=& G(C_h S_w + C_i S_h C_w  ) - \frac{5}{32} \sqrt{\frac{21}{\pi}} C_i S_i^2 S_h \big(4 S_w S_{2w}   f_{3,1,3} 
                    +  C_w  T_3  \big)
                    \nonumber \\
                    &-& \frac{3}{4} S_i^2 \bigg( \sqrt{\frac{5}{\pi}} S_{2w} f_{2,0,2} - \frac{5}{8} \sqrt{\frac{21}{\pi}} C_h S_{3w} f_{3,1,3} + \frac{5}{24} \sqrt{\frac{21}{\pi}} C_h S_w f_{3,1,1}              \bigg)
\\
L_p \dot{\alpha}_p &=& G(S_h C_w + C_i C_h S_w  )  + \frac{5}{32} \sqrt{\frac{21}{\pi}} S_i^2  ( S_h 
T_2  - C_i C_h S_w T_3         )
\\
L_p \dot{\omega}_p &=& \frac{\Gamma}{6 l_p^4} (15 C_i^2 - 3) + G l_p^{-1} C_i S_h S_w + \frac{5}{32} \sqrt{\frac{21}{\pi}} l_p^{-1} C_i (3 C_i^2 - 1) S_h S_w T_3  + l_p^{-1} C_i^2 \bigg[\frac{3}{4} \sqrt{\frac{5}{\pi}} T_1 + \frac{5}{16} \sqrt{\frac{21}{\pi}} C_h T_2                \bigg]
\nonumber \\
& -  &  \frac{l_p}{\sqrt{1-l_p^2}}  \bigg[ F^{'} 
+ G^{'} ( C_h C_w - C_i S_h S_w) + S_i^2 \bigg(    \frac{3}{8} \sqrt{\frac{5}{\pi}} T_1^{'} + \frac{5}{32} \sqrt{\frac{21}{\pi}} C_h T_2^{'}        \bigg) 
+ \frac{5}{32} \sqrt{\frac{21}{\pi}} S_h S_w C_i S_i^2 T_3^{'} \bigg]
\label{eq:wpdot_3D}
\end{eqnarray}
where we have written
\begin{subequations}
\begin{align}
& L_p = \sqrt{G M_\odot a_p} \\
& T_1 = f_{2,0,0} - C_{2w} f_{2,0,2} \\
& T_2 = C_{3w} f_{3,1,3} - C_{w} f_{3,1,1} \\
& T_3 = 3 f_{3,1,1} - f_{3,1,3} ( 1 + 2 C_{2w} )
\end{align}
\end{subequations}
and $S$ and $C$ are shorthand for sine and cosine of the angles given as subscript: $i$ is the inclination determined by $\alpha \equiv L_z / L_p = l_p \cos(i_p)$, $l_p = \sqrt{1-e_p^2}$ as before, $h = \Omega_p - \varpi_d$ and $w$ is argument of pericenter. Note that the primed terms in Eq. \ref{eq:wpdot_3D} are defined such that $f^{'} = \frac{\partial  f(e_p) }{\partial e_p} $.
This set of equations represents the basis of our population study performed in the body of this work (Section \ref{section:3}).

Finally, we comment on the accuracy of our numerical disk potential. Insisting on coplanar test-particle orbits by setting $i_p = 0$ in Eq. \ref{eqn:planets_H3D}, we solved for the equilibria and analyzed their stability. The results are shown in Fig. \ref{fig:fig15}. It is evident that the coplanar equilibria recovered by the numerical formulation of the disk potential (Eq. \ref{eqn:numerical_H3D}) and that of the analytical (Eq. \ref{eqn:Rd}) agree very well both in orbital structure and stability.
%

\end{document}